# Title: Ultrahigh Anomalous Nernst Thermopower and Thermal Hall Angle in YbMnBi$_2$

**Authors:** Jiamin Wen[1], Kaustuv Manna[2,3], Dung Vu[4,5]†, Subhadeep Bej[3], Yu Pan[2,6]‡, Claudia Felser[2], Brian Skinner[7]#, Joseph P. Heremans[1,4,7]#

**Affiliations:**

[1]Department of Materials Science and Engineering, The Ohio State University; Columbus, 43210, USA.

[2]Max Planck Institute for Chemical Physics of Solids; Dresden, 01187, Germany.

[3]Department of Physics, Indian Institute of Technology Delhi; New Delhi, 110016, India.

[4]Department of Mechanical and Aerospace Engineering, The Ohio State University; Columbus, 43210, USA.

[5]Department of Applied Physics, Yale University; New Haven, 06520, USA.

[6]College of Materials College of Materials Science and Engineering and Center of Quantum Materials & Devices, Chongqing University, Chongqing, China.

[7]Department of Physics, The Ohio State University; Columbus, 43210, USA.

†The present address of this author is [5].

‡The present address of this author is [6].

#Corresponding authors. Email: skinner.352@osu.edu, heremans.1@osu.edu

**Abstract:** Thermoelectrics (TEs) are solid-state devices that can realize heat-electricity conversion. Transverse TEs require materials with a large Nernst effect, which typically requires a strong applied magnetic field. However, topological materials with magnetic order offer an alternative pathway for achieving large Nernst via the anomalous Hall effect and the accompanying anomalous Nernst effect (ANE) that arise from band topology. Here, we show that YbMnBi$_2$ with a low Hall density and a chemical potential near the Weyl points has, to the best of our knowledge, the highest ANE-dominated Nernst thermopower of any magnetic material, with $S_{yx}$ around 110 μV K$^{-1}$ ($T$ = 254 K, 5 T ≤ |$μ_0H$| ≤ 9 T applied along the spin canting direction), due to the synergism between classical contributions from filled electron bands, large Hall conductivity of topological origin, and large resistivity anisotropy. An appreciable thermal Hall angle of $0.02 < \nabla_y T/\nabla_x T$ (-9 T) $< 0.06$ was observed (40 K $< T <$ 310 K).

**Introduction:**

Thermoelectric (TE) power generation and refrigeration are enabling technologies for climate change abatement, providing means to convert otherwise wasted heat into useful electrical energy and greenhouse-gas-free cooling and refrigeration[1-4]. So far, extensive thermoelectric research has been focusing on the longitudinal effect, i.e., Seebeck effect, in which the generated potential gradient is parallel to the applied heat flux. TE modules based on this effect are piles made from multiple thermocouples, each made of separate materials with different polarities of charge carriers. This construction is necessary to build up useful voltages[1], but it requires multiple metal/semiconductor electrical connections, each with an unavoidable electrical and thermal contact resistance leading to the loss of efficiency. In contrast, transverse TEs utilizing Nernst effect, in which the created voltage is perpendicular to the applied temperature gradient, may vastly simplify the module fabrication[2]. They require only one material, can be optimized by changing the geometry only, and make it possible to set up electrical contacts only at one end, preferably the one near room temperature. These advantages circumvent the need of longitudinal devices to have contacts at the hot side of thermoelectric generators (TEGs), where they degrade. In TE coolers, single-stage longitudinal (Peltier) coolers can reach only a limited temperature difference, so that large temperature drops require complex and delicate multi-stage devices. Transverse (Nernst-Ettingshausen) TE coolers bypass this limitation as well. In the present work, we demonstrate how harnessing the anomalous Nernst effect (ANE) in a magnetic Weyl semimetal (WSM) contributes to a record transverse thermopower.

In recent years, more and more efforts have been dedicated to the exploration of ANE[5-17], aiming at considerable transverse thermoelectric performance. While nonmagnetic semimetals with high mobility and low carrier concentration can produce a large Nernst effect in a strong applied magnetic field (e.g., Refs. 18-19), the ANE offers the promise of a strong Nernst response without requiring a strong applied field, since the time reversal symmetry breaking required for a transverse response is already intrinsic to the material. ANE has been commonly observed in ferromagnets (FMs) [5,8-11,13,15,16]. $Co_2MnGa$[8,9], $Co_3Sn_2S_2$[10,11], and $Fe_3Ga$[13] exhibited ANE coefficients ($S_{ANE}$) ranging from 3 to 8 µV K$^{-1}$. In addition, $UCo_{0.8}Ru_{0.2}Al$[15] was found to possess an $S_{ANE}$ of 23 µV K$^{-1}$, a record at that time. FMs can break time-reversal symmetry (TRS) and give rise to a non-zero Berry curvature, which is the intrinsic origin of ANE[7,8,13,20], but the disadvantages of FMs in thermoelectric applications are also obvious: 1. they usually exhibit high electrical resistivity resulting from heavy band mass; 2. they are prone to create magnetic disruption owing to the high magnetization and strong inherent stray field[17].

Given the difficulties described above, searching for other materials retaining the advantages of FMs while overcoming their drawbacks has been a pressing challenge. In 2022, Pan et al. reported that $YbMnBi_2$, an antiferromagnetic (AFM) WSM with spins canted in the [110] direction providing broken TRS[21], has low resistivity arising from the light effective band mass, weak magnetization, a substantial ANE thermopower $S_{ANE}$ of up to 6 µV K$^{-1}$ and ANE conductivity $\alpha_{ANE}$ of up to 10 A m$^{-1}$ K$^{-1}$ in the configuration of $H$ // [100], $j_Q$ // [010], $V$ // [001] and $H$ // [100], $j_Q$ // [001], $V$ // [010], respectively[17]. These responses were claimed to transcend all the previously reported large $S_{ANE}$ and $\alpha_{ANE}$ values in AFMs[6,12,14,22], making $YbMnBi_2$ an encouraging candidate for Nernst devices (the properties of those samples are summarized in Supplementary Table 1). The Nernst thermopower $S_{yx}$ is the ratio of the transverse electric field to the longitudinal temperature gradient, whereas the Nernst conductivity $\alpha_{yx}$ is the ratio of the transverse current flow to the longitudinal temperature gradient; both are related as will be shown later.

Because the results depend critically on the position of the chemical potential vis-à-vis the Weyl points, and because this potential depends on unintentional doping of the material, those early transport experiments on this material were not yet comprehensive enough to unveil the potential of YbMnBi$_2$. In this report, we examine two YbMnBi$_2$ samples (sample 1 and sample 2) with lower Hall densities (2.22 × 10$^{19}$ cm$^{-3}$ at $T$ = 8 K for sample 1) and higher Hall mobilities (1.05 × 10$^5$ cm$^2$ V$^{-1}$ s$^{-1}$ at $T$ = 8 K for sample 1) than the samples in ref. 17, bringing the chemical potential $\mu$ closer to the Weyl points. We show here how this change in $\mu$ drastically changes the physical origin of $S_{\mathrm{ANE}}$. In ref. 17 with heavier-doped material, the Nernst response is dominated by $\alpha_{\mathrm{ANE}}$ with significant contributions from some indefinite extrinsic sources. Here, in contrast, the record value of $S_{\mathrm{ANE}}$ is due to the synergistic effect between a topological anomalous Hall conductivity, a Nernst conductivity that is non-topological in origin, and the strong anisotropy of the dispersion of the Weyl bands near the Weyl points. Supplementary Table 1 compares the properties of samples 1 and 2 used here and those in ref. 17; Supplementary Fig. 3 gives the temperature dependence of our Hall density and mobility. We caution that the interpretation of the Hall density and its temperature dependence are complicated by the coexistence of multiple bands at the Fermi level. We discuss this issue in detail in the Supplementary Information.

Another key difference with the previous work of Pan et al. is that in their study the external magnetic field, applied temperature gradient, and measured voltage signal were along the three primitive axes. Here, we performed transport property measurements with field applied along the spin canting direction [110] (Fig. 1c, 1d, 1e, and 1f). We report that sample 1 in this study has a Nernst thermopower that sets a significantly enhanced record in magnetic materials, now reaching maximum $S_{yx}$ ~ 110 μV K$^{-1}$ in the configuration of $H$ // [110], $j_Q$ // [1$\bar{1}$0], and $V$ // [001], likely ANE-dominated (see Supplementary Table 1), and a considerable thermal Hall effect (THE) angle $0.02 < \nabla_y T/\nabla_x T$ (-9 T) $< 0.06$. In the Supplementary, we present results on sample 2, which has a higher Hall density and lower Hall mobility than sample 1, implying its chemical potential is farther from the Weyl points. Qualitatively, the results of this study are notably reproducible, although the Nernst coefficient (maximum $S_{yx} \sim 38$ µV K$^{-1}$) is smaller in sample 2 than in sample 1.

To the best of our knowledge, these lower-doped YbMnBi$_2$ samples have the highest anomalous Nernst thermopower of any magnetic materials (see Table 1). Below we present evidence that the synergism between the topological nature of the Weyl bands, classical effects in filled bands, and anisotropy plays a pivotal role in producing this large Nernst thermopower. Together our results provide robust evidence for a large topological contribution to the Nernst effect that is driven by a spin canting transition, and in this way point toward a new pathway for designing efficient transverse thermoelectrics by exploiting band topology arising from field-tunable magnetic order.

**Results and Discussion**

As shown by Fig. 1a, the crystal structure of YbMnBi$_2$ falls into a tetragonal system, whose space group is *P4/nmm*. There are two types of Bi atoms: type I forms nearest-neighbors of Mn atoms while type II constitutes an interlayer. The Néel temperature $T_N$ is reported to be around 290 K and the antiferromagnetic ordering of Mn atoms with spins aligned along *c* axis within about 3° is confirmed[24]. The spin canting direction is [110]. Pan et al. published that the spin canting temperature (i.e., below which the spin canting shows up) is ~ 250 K with a small canting angle[17]. Although the magnitude of spin canting is minor, it induces the formation of Weyl points which act as sources of Berry curvature, as we describe below, and these Weyl points allow for an anomalous Hall response.

YbMnBi$_2$ contains topological bands arising primarily from $p_{x/y}$ orbitals on the interlayer Bi atoms[17,21]; as is obvious from the nearly two-dimensional arrangement of these Bi atoms, these bands are extremely anisotropic, an important attribute for the observations reported here. If one uses a description that assumes simple antiferromagnetic ordering of the Mn spins and neglects spin-orbit coupling, then the topological bands form a nodal line slightly below the Fermi level. When spin-orbit effects are included, the nodal line becomes gapped everywhere except at four Dirac points. The effect of spin canting is to split each of these four Dirac points into two Weyl points with opposite chirality, leaving 8 Weyl points in total[17,21]. The splitting of the Dirac points into Weyl point pairs allows for the possibility of an anomalous Hall conductivity that is proportional to the separation in momentum space of the two opposite-chirality Weyl points within a pair[25,26].

The resulting linear dispersion near the Weyl points has a very large Fermi velocity ($\approx 1 \times 10^6$ m/s) within the *ab* plane, and a much smaller Fermi velocity ($\approx 6 \times 10^3$ m/s) in the *c* (layering) direction[21,27]. Thus, the Fermi surface in the vicinity of the Weyl points has a highly elongated cigar shape, stretched in the *c* direction (as depicted in Fig. 1b). Coexisting with the Weyl bands is a heavy, trivial hole band associated primarily with Mn *d* orbitals[17].

*Giant Anomalous Nernst Effect Thermopower*

Figure 2a shows the behavior of the Nernst coefficient $S_{yx}$ with respect to magnetic field at a variety of temperatures for sample 1 (the data on sample 2 are in Supplementary Fig. 4a). Consistent with each other, $S_{yx}$ of both sample 1 and sample 2 first increases and then decreases with temperature when 5 T $\leq |\mu_0 H| \leq$ 9 T, achieving a maximum around 250 K. This nonmonotonic dependence of the Nernst effect on temperature is characteristic of nodal

semimetals with low Fermi energy, for which thermally excited holes in the valence band cause the Hall conductivity to drop when $k_BT$ becomes comparable to or larger than $E_F$[28,29].

Both samples exhibit a characteristic combination of ONE (ordinary Nernst effect) and ANE. We caution that it is difficult in general to separate the contributions of ONE and ANE to the total Nernst effect. The usual method used in literature is to assume that the ONE is linear in magnetic field and to subtract this linear dependence from the total $S_{yx}$ in order to estimate the ANE. However, the higher the carrier mobility, the narrower the linear regime of the field dependence of ONE, and when the ONE is beyond the linear region the superimposition of ONE and ANE has complicated features, and this linear subtraction becomes questionable. In the Supplementary Fig. 4b, $S_{ANE}$ of sample 2 is tentatively derived from the data by assuming the ONE contribution is linear and then subtracting it out following the procedures outlined in ref. 17. The resulting estimated maximum value in sample 2 is $S_{ANE} \sim 30$ μV K$^{-1}$ when 5 T $\leq |\mu_0H| \leq$ 9 T and 200 K < T < 260 K; Apparently, ANE dominates $S_{yx}$ of sample 2, and this is a record-high ANE thermopower as well, exceeded only by sample 1. Regarding sample 1, at $T = 254$ K, $S_{yx}$ goes up to ~ 110 μV K$^{-1}$ when 5 T $\leq |\mu_0H| \leq$ 9 T and stays almost constant there, also presumably with $S_{ANE}$ dominant.

Figures 2 and 3 give the transverse and longitudinal electronic transport properties respectively for sample 1 (Supplementary Figs. 4 and 6 for sample 2). The Hall resistivity $\rho_{yx}$ of sample 1 is given in Fig. 2b. The in-plane and cross-plane resistivities $\rho_{xx}$ and $\rho_{yy}$ are in Fig. 3a and 3d, respectively. $\rho_{yy}$ increases as temperature rises and its values are surprisingly large. The anisotropy ratio of the longitudinal resistivity $\rho_{yy}/\rho_{xx}$ observed here is at the level of thousands, much larger than that in the prior study (which saw $\rho_{yy}/\rho_{xx}$ up to 30 (ref. 17)). This large anisotropy is accounted for by the anisotropic nature of the Fermi surface, which is highly dispersive in the xz plane but has little dispersion along y axis[30]. Our measured values of $\rho_{yy}$ are dozens of times higher than the values reported in ref. 17, while the in-plane resistivity is smaller. Both observations are attributed to the fact that the chemical potential in the present samples is closer to the Weyl points. What's more, $\rho_{yy}$ of sample 2 (Supplementary Fig. 6d) unexpectedly showcases a negative magnetoresistance (NMR) of up to ~ 25% at low temperatures, which persists through the whole range of measured temperatures (47 K ~ 307 K), indicating a temperature-robust NMR, while only a slight positive magnetoresistance is observed in $\rho_{yy}$ of sample 1 (Fig. 3d).

From these data, the Hall conductivity $\sigma_{yx} = -\rho_{yx}/(\rho_{xx}\rho_{yy} + \rho_{yx}^2)$ can be derived (see Supplementary Information) and is given in Fig. 2c for sample 1 (Supplementary Fig. 4d for sample 2). $\sigma_{yx}$ exhibits a jump as a function of magnetic field $H$ near $H = 0$, and this jump becomes increasingly sharp as the temperature is lowered. Such a jump is consistent with an anomalous Hall effect (AHE) arising from the filled Weyl band, with the sign of the Berry curvature at the Weyl points becoming inverted when the magnetic field direction is inverted and switches the direction of spin canting. The in-plane thermopower $S_{xx}$ and thermal conductivity $\kappa_{xx}$ are given in Fig. 3b and 3c, respectively.

This behavior of $\sigma_{yx}$ can be contrasted with the behavior of the Nernst conductivity, which is given by (see Supplementary Information):

$$\alpha_{yx} = \frac{-\rho_{yx}S_{xx} + \rho_{xx}S_{yx}}{\rho_{xx}\rho_{yy} + \rho_{yx}^2} \qquad (1)$$

where $S_{xx}$ and $S_{yx}$ are the Seebeck and Nernst coefficients, respectively. The experimental value of $\alpha_{yx}$ is plotted in Fig. 2d for sample 1 (Supplementary Fig. 4e for sample 2) as a function of magnetic field at various temperatures. The anomalous Nernst conductivity $\alpha_{ANE}$ for sample 2 (Supplementary Fig. 4g) can be estimated following the same method as for estimating $S_{ANE}$. An explicit and quantitative explanation for $\alpha_{ANE}$ was attempted in ref. 17, but this explanation did not explain the temperature dependence well even in highly doped samples, which made the authors invoke the possibility that extrinsic mechanisms contributing to the ANE might also be present, such as skew scattering[31,32] and side jump scattering[33,34]. Interestingly, while ref. 17 showed trends of $T$-dependence of $\alpha_{ANE}$ that were opposite from theoretical predictions, in this study we find an $\alpha_{ANE}$ with a $T$-dependence that is generally consistent with the first-principle calculations presented in ref. 17, as shown in Supplementary Fig. 4h. This consistency allows us to explain our data with a relatively simple analytical model based on Weyl/Dirac semimetal physics (see Supplementary Information for a detailed discussion); the calculated $\alpha_{yx}$ is given in Fig. 2e for sample 1 (Supplementary Fig. 4f for sample 2) and describes the data quite well. Unlike $\sigma_{yx}$, which has a constant contribution from the filled bands, $\alpha_{yx}$ is proportional to $\frac{d\sigma_{yx}}{dE}|_{E=E_F}$ (ref. 35, see Supplementary Information for a detailed discussion), so that the filled bands do not contribute. Consequently, in our samples $\alpha_{yx}$ is smaller in magnitude than that in ref. 17, does not exhibit a sharp jump around $H = 0$, and is consistent with a theoretical description that does not invoke Berry curvature. Our theoretical description of $\alpha_{yx}$ contains two parameters which depend on the Fermi energy and mobility. The fitted values of these parameters are consistent with sample 1 having a lower Fermi energy and higher mobility than sample 2, as we describe in detail in the Supplementary Information.

Why, if the Nernst conductivity $\alpha_{yx}$ is smaller than that in ref. 17, is the Nernst thermopower that much larger? As mentioned above, the main difference between the samples in this study and in ref. 17 are the much lower Hall densities and much higher Hall mobilities, due to changes in the unintentional doping levels. This brings the chemical potential closer to the Weyl points, as evidenced by two observations: (i) the Hall density increases with temperature ($T > 80$ K) (shown in Supplementary Fig. 3) and (ii) the polarity of high-field $d\rho_{yx}/dH$ flips as temperature changes (shown in Fig. 2b). Both behaviors are characteristic of intrinsic semiconductors or semimetals with small or zero band overlap: they imply the dominance of thermally excited intrinsic charge carriers. (We discuss the behavior of the Hall resistivity in more detail in the Supplementary Information.)

Rewriting Eq. (1), it now is possible to back-calculate $S_{yx}$ from:

$$S_{yx} = \frac{\alpha_{yx} - \sigma_{yx} S_{xx}}{\sigma_{yy}} \qquad (2)$$

The derivation of Eq. (2) is in the Supplementary Information. The back-calculated values of $S_{yx}$, which fit the data very well, are given in Supplementary Fig. 5. Eq. (2) gives insight into the origin of the record $S_{yx}$. Observing that $\alpha_{yx}$ and the product $\sigma_{yx}S_{xx}$ have opposite signs, the large numerator of this equation is a result of the fact that the absolute values of the two quantities add up, producing a sum of the absolute values of a topological term ($\sigma_{yx}$) and a classical one ($\alpha_{yx}$). The Nernst thermopower is then further boosted by the large anisotropy of the resistivity, producing a small denominator. It can thus be concluded that the record $S_{yx}$ results from the synergy of anisotropy, classical and topological transport.

*Appreciable Thermal Hall Effect Angle*

The thermal Hall effect (THE) is another transport property that is strongly influenced by the Berry curvature of the carriers. Fig. 4a shows the field dependence of the thermal Hall resistivity $w_{xy}$ of sample 1 at varying temperatures, which is relatively large, up to ~ $3.3 \times 10^{-3}$ W$^{-1}$ m K. The data for sample 2 (Supplementary Fig. 7a) exhibit excellent consistency. In YbMnBi$_2$, $w_{xy}$ behaves approximately linearly with field at low temperatures ($T < 100$ K for samples 1 and 2), displays a plateau-like, field-independent character in the intermediate range (100 K < $T$ < 250 K for samples 1 and 2), and then tends to be linear in field again as temperature further increases. The THE is of electronic origin and thus largely reflects the electrical Hall effect, which has a substantial contribution from the Berry curvature of the Weyl bands. While a similarly large or even larger THE signal can appear in trivial materials if the mobility is high and the carrier density and lattice thermal conductivity are sufficiently low[36-38], the sharp step in $w_{xy}$ near $B = 0$ is qualitatively different from what appears in that context and suggests that there is a large anomalous contribution to the THE in addition to the trivial contribution.

In THE measurements, the transverse temperature difference is measured for a given applied heat flux, meaning that the actually measured variable is one of the thermal Hall resistivity terms in the thermal resistivity matrix. The commonly used representative for THE is the so-called thermal Hall conductivity $\kappa_{xy}$ calculated through the inversion of the thermal resistivity matrix. Here, the corresponding formula can be written as $\kappa_{xy} = -w_{xy}/(w_{xx}w_{yy} + w_{xy}^2)$, where $w_{xx}$ and $w_{yy}$ are thermal resistivities along $x$ and $y$ direction, respectively. In general, $w_{xx}w_{yy} \gg w_{xy}^2$, so $\kappa_{xy}$ approximates $-w_{xy}/w_{xx}w_{yy}$. If either $w_{xx}$ or $w_{yy}$ depends strongly on magnetic field, the longitudinal thermal magnetoresistance will heavily contaminate the dependence of thermal Hall signal on magnetic field. Considering this weakness, to get a clear picture of THE in YbMnBi$_2$, the tangent of thermal Hall angle $\tan(\theta) \equiv \nabla_y T/\nabla_x T = w_{xy}/w_{xx}$ is reported in Fig. 4c for sample 1 and Supplementary Fig. 7c for sample 2: it simply equals the ratio of the transverse and longitudinal components of the temperature gradient, does not require matrix operation, and incorporates the magnetoresistance in the longitudinal thermal transport. The field dependence and temperature dependence of $\nabla_y T/\nabla_x T$ is analogous to those of $w_{xy}$. The absolute values of $\nabla_y T/\nabla_x T$ at individual magnetic fields do not change rapidly with temperature, supported by the observation that $0.02 < \nabla_y T / \nabla_x T$ (-9 T) $< 0.06$ while $T$ changes between 40 K and 310 K.

In summary, we report in the magnetic Weyl semimetal YbMnBi$_2$ the largest anomalous Nernst thermopower observed in a magnetic solid and attribute it to the synergism of a Berry-curvature-induced Hall conductivity, a classical Nernst conductivity, and a highly anisotropic resistivity. This synergy works only if the carrier density is low and the chemical potential is close to the Weyl points. The samples also have a sizable thermal Hall effect, benefiting from the significant topological contributions of the Weyl bands. These results open up the possibility of furthering the leading edge of research on transverse (Nernst) thermoelectrics and provide a powerful pathway to do so, by means of Fermi level tuning in materials with exotic band topology and structures.

41. Xu, L. C. et al. Anomalous transverse response of $Co_2MnGa$ and universality of the room-temperature $\alpha_{ij}^A/\sigma_{ij}^A$ ratio across topological magnets. *Phys. Rev. B* **101**, 180404(R) (2020).
16

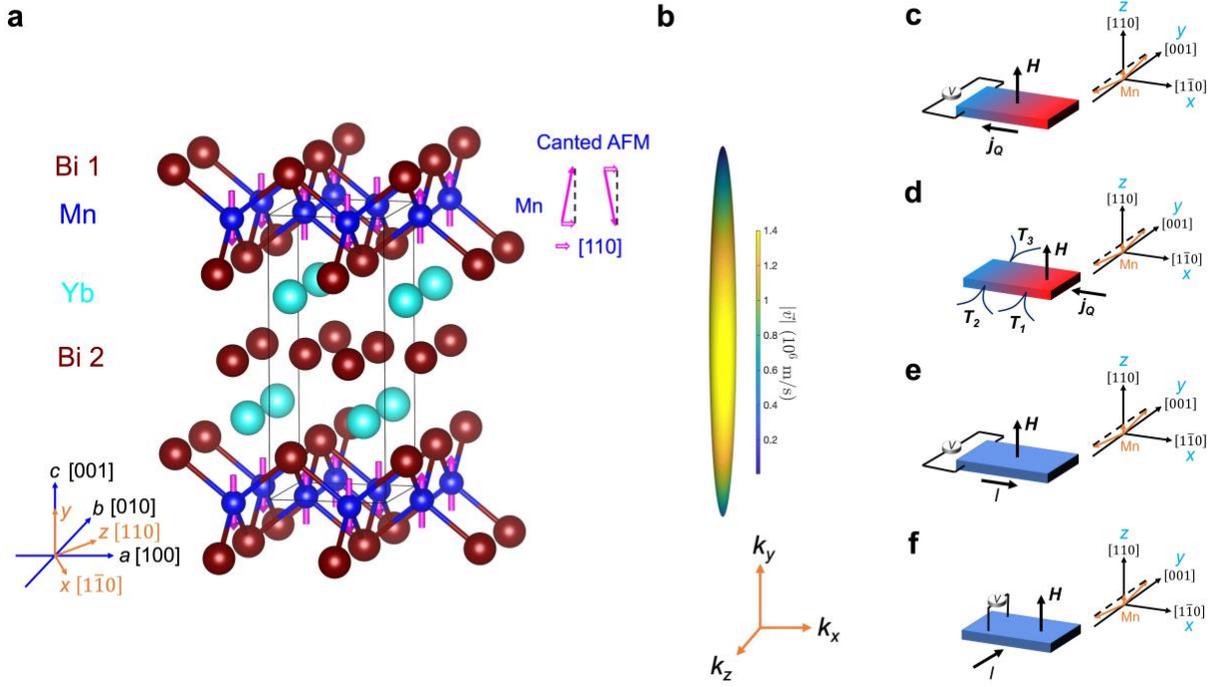

**Fig. 1 | Crystal structure of YbMnBi$_2$, schematic Fermi surface, and experimental setups. a**, The spins of Mn atoms in single crystal YbMnBi$_2$ are antiparallel within the *ab* plane, slightly canted along the [110] direction, and ferromagnetically stacked along the *c* axis. **b.** Schematic depiction of the Fermi surface near the Weyl points. The Fermi surface is extremely elongated along the *c* direction, while the Fermi velocity (indicated by the color) is much larger in the *ab* plane. Note that this image is broadened in the $k_x$-$k_z$ plane for visual clarity; the actual aspect ratio of the Fermi surface is closer to 200:1. **c**, A schematic drawing of the experimental configuration of the Nernst effect measurement, in which magnetic field *H* was applied along the spin canting direction [110], heat flux $j_Q$ was along [1$\bar{1}$0], and recorded transverse voltage *V* was along [001]. **d**, A schematic drawing of the experimental configuration of the thermal Hall measurement, in which magnetic field *H* was applied along the spin canting direction [110], heat flux $j_Q$ was along [1$\bar{1}$0], and the transverse temperature difference $\Delta_y T$ was obtained through the subtraction of the readouts of the two transverse thermometers, which is $T_3 - T_2$ in this setup. **e**, A schematic drawing of the experimental setup of the electrical Hall experiment, in which external magnetic field *H* was along the spin canting direction [110], applied current flow *I* was along [1$\bar{1}$0], and measured transverse voltage *V* was along [001]. **f**, A schematic drawing of the experimental setup of a separate electrical transport experiment, in which external magnetic field *H* was along the spin canting direction [110], applied current flow *I* was along [001], and measured voltage *V* was also along [001].

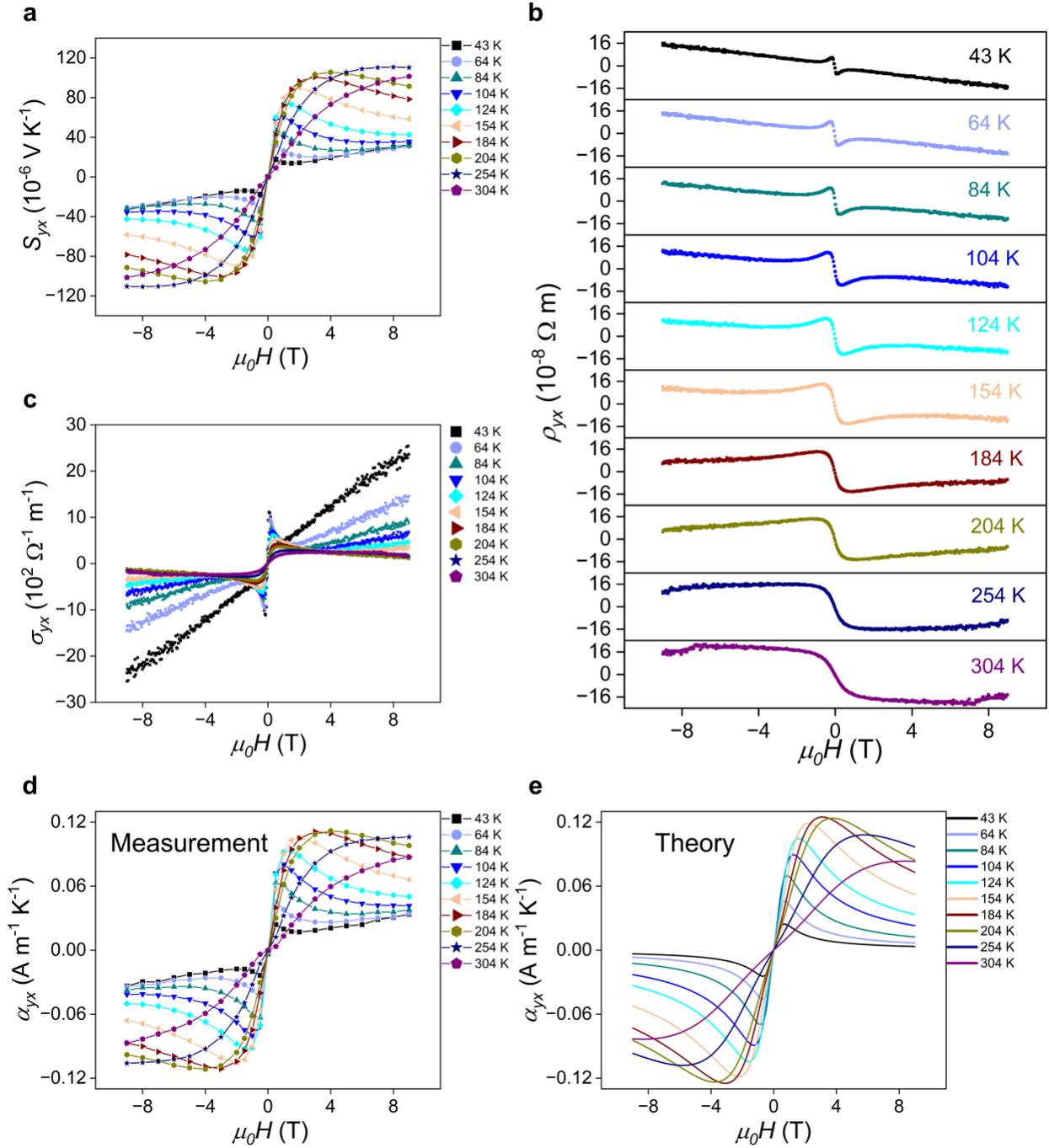

**Fig. 2 | Transverse transport properties of sample 1. a**, Field dependence of Nernst thermopower $S_{yx}$ at various temperatures. The similarity of the results of this sample with those on sample 2 (see Supplementary Fig. 4a) confirms that the observation of the enormous Nernst thermopower is reproducible. **b**–**e**, Field dependence of (**b**) electrical Hall resistivity $\rho_{yx}$, (**c**) electrical Hall conductivity $\sigma_{yx}$, (**d**) Nernst conductivity $\alpha_{yx}$ obtained from measured properties outlined by Eq. (1), and (**e**) theoretical curves for the Nernst conductivity $\alpha_{yx}$, at various temperatures.

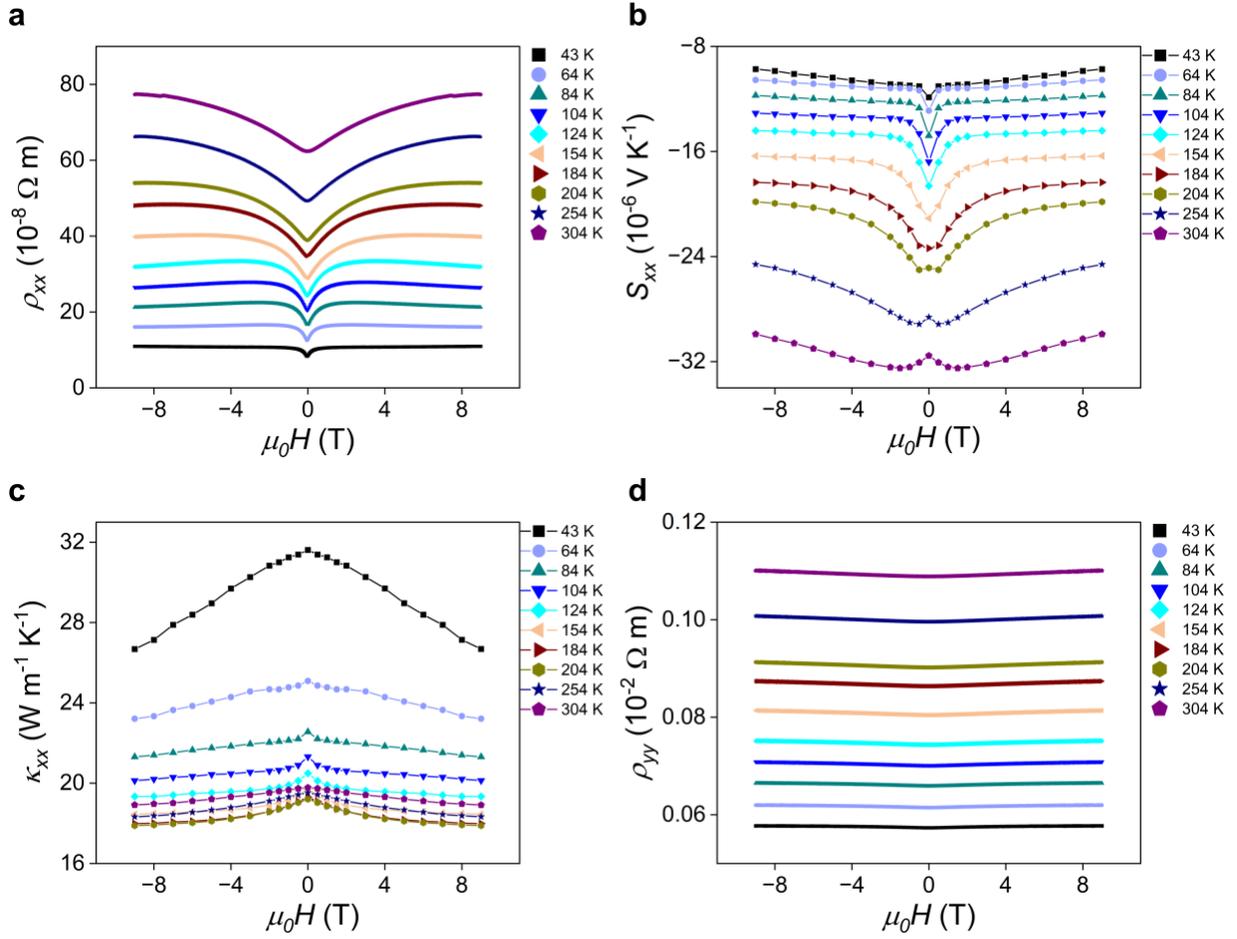

**Fig. 3 | Longitudinal transport properties of sample 1. a–d**, Field dependence of (**a**) electrical resistivity $\rho_{xx}$, (**b**) Seebeck coefficient $S_{xx}$, (**c**) thermal conductivity $\kappa_{xx}$, and (**d**) electrical resistivity $\rho_{yy}$, at various temperatures.

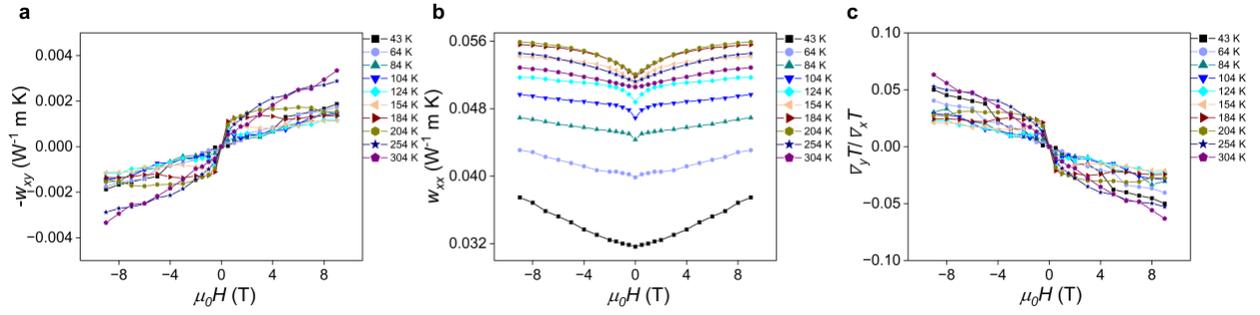

**Fig. 4 | Appreciable thermal Hall effect signal (sample 1). a–c**, Field dependence of (**a**) thermal Hall resistivity $w_{xy}$, (**b**) thermal resistivity $w_{xx}$, and (**c**) the tangent of thermal Hall angle $\nabla_y T/\nabla_x T$, at various temperatures.

| Compound | Max $S_{ANE}$ (µV K$^{-1}$) |
|---|---|
| YbMnBi$_2$ (sample 1 this work) | > 70 |
| YbMnBi$_2$ (sample 2 this work) | ~ 30 |
| YbMnBi$_2$ (ref. 17) | ~ 6 |
| MnBi (ref. 16) | ~ 10 |
| CeCrGe$_3$ (ref. 23) | ~ 10 |
| Co$_2$MnGa (ref. 8) | ~ 8 |
| Fe$_3$Ga (ref. 13) | ~ 6 |
| UCo$_{0.8}$Ru$_{0.2}$Al (ref. 15) | ~ 23 |

**Table 1 | Comparison of ANE thermopower $S_{ANE}$ of this study with some of the largest reported $S_{ANE}$ values in AFMs and FMs.** The previously documented largest $S_{ANE}$ in AFMs is ~ 6 µV K$^{-1}$ in the same compound as this work but in a different configuration. The other referenced compounds in the table are FMs, among which UCo$_{0.8}$Ru$_{0.2}$Al established the apex $S_{ANE}$ value of ~ 23 µV K$^{-1}$. Even though for sample 1 of this study, we cannot reliably separate the ANE from the total Nernst thermopower, the tendency of values of $S_{yx}$ above 180 K is to saturate above 70 µV K$^{-1}$; therefore, the findings of the present research may well set a new world record for ANE thermopower in magnetic materials.

## Methods

Sample preparation

The single crystals of YbMnBi$_2$ measured in this research were synthesized at Max Plank Institute for Chemical Physics of Solids (CPfS) using the self-flux method with an element molar ratio of Yb:Mn:Bi = 1:1:10. Yb of purity 99.99%, Mn of purity 99.98%, and Bi of purity 99.999% were pruned into small pieces, mixed up, and then placed in an alumina crucible. The mixture was sealed in a quartz tube under a partial argon pressure. The temperature of the sealed tube was increased to 1273 K with a rate of 50 K/h, held there for 20 hours, and then cooled down slowly to 673 K at a rate of 3 K/h. Finally, single crystals of YbMnBi$_2$ were obtained by removing the Bi flux through centrifugation at 673 K. Given that YbMnBi$_2$ is a relative new material system and a ternary compound, control over its defect chemistry is not yet well developed. Unintentional doping occurs. A study of the dependence of the transport properties on $\mu$ is therefore carried out by selecting and characterizing different crystals from the melt rather than by deliberate extrinsic doping, as can be done for heritage semiconductors and semimetals. Two pieces of single crystals, labeled as Sample 1 and Sample 2, respectively, were used throughout this work to examine the role of $\mu$ and compare the results to those of the previous work[17].

Transport property and magnetization measurements

Both thermal transport and electrical transport properties were measured in a Quantum Design PPMS® (Physical Property Measurement System; serial number: PPMS149) with a breakout box via a standard four-probe steady-state method[39]. National Instruments™ LabVIEW was utilized to build the customized test protocols and control the associated instruments. The magnetization measurements were conducted in a SQUID (Superconducting Quantum Interference Device) magnetometer, which specifically is Quantum Design MPMS®3.

Sample characterization

Laue X-ray diffraction was employed to determine the single-crystallinity and orientation of the as-grown crystals. Energy-dispersive X-ray spectroscopy (Quantax, Bruker) was utilized to examine the composition. Combining these two methods, with the benefit of the non-destructive nature of the instrumental approach, the phase purity of the as-grown samples can be confidently verified. X-ray diffraction (XRD) test on an after-measurement mounted sample was conducted by using Bruker D2 PHASER.

**Data availability**

Data supporting the findings of this study are available within the paper or from the authors.

**Acknowledgments:** J.W., D.V., B.S., and J.P.H. acknowledge the funding support from the National Science Foundation grant DMR 2011876 "Center for Emergent Materials", an NSF MRSEC. K.M., Y.P., and C.F. acknowledge the funding support from Deutsche Forschungsgemeinschaft (DFG) under SFB1143 (Project No. 247310070), the Wuerzburg-Dresden Cluster of Excellence on Complexity and Topology in Quantum Matter – ct.qmat EXC 2147, Project No. 390858490. K.M. and S.B. acknowledge Max Planck Society for funding support under the Max Planck–India partner group project and the Central Research Facility (CRF), IIT Delhi, for providing characterization facility FESEM-EDX to carry out compositional analysis with elemental mapping. We thank S. J. Watzman for helpful discussions.

**Author contributions:**

Conceptualization: J.P.H., J.W., D.V., Y.P.

Methodology: J.P.H., J.W., D.V., Y.P.

Experimental Investigation: J.W., S.B., D.V., K.M., Y.P.

Modeling: B.S., J.W., J.P.H.

Funding acquisition: J.P.H., B.S., C.F.

Project administration: J.P.H., C.F.

Supervision: J.P.H., C.F., Y.P.

Writing – original draft: J.W., B.S., J.P.H.

Writing – review & editing: All

# Supplementary Information

# for

## Ultrahigh Anomalous Nernst Thermopower and Thermal Hall Angle in YbMnBi$_2$

**Authors:** Jiamin Wen[1], Kaustuv Manna[2,3], Dung Vu[4,5]†, Subhadeep Bej[3], Yu Pan[2,6]‡, Claudia Felser[2], Brian Skinner[7]#, Joseph P. Heremans[1,4,7]#

**Affiliations:**

[1]Department of Materials Science and Engineering, The Ohio State University; Columbus, 43210, USA.

[2]Max Planck Institute for Chemical Physics of Solids; Dresden, 01187, Germany.

[3]Department of Physics, Indian Institute of Technology Delhi; New Delhi, 110016, India.

[4]Department of Mechanical and Aerospace Engineering, The Ohio State University; Columbus, 43210, USA.

[5]Department of Applied Physics, Yale University; New Haven, 06520, USA.

[6] College of Materials College of Materials Science and Engineering and Center of Quantum Materials & Devices, Chongqing University, Chongqing, China.

[7]Department of Physics, The Ohio State University; Columbus, 43210, USA.

†The present address of this author is [5].

‡The present address of this author is [6].

#Corresponding authors. Email: skinner.352@osu.edu, heremans.1@osu.edu

**Confirmation of nonexistence of bismuth or MnBi inclusions**

As demonstrated by Supplementary Fig. 1e, no peaks from the XRD data of the after-measurement sample 1, which was mounted as a device, reasonably match the calculated powder XRD pattern of elemental bismuth and MnBi compound. This confirms that there is no elemental bismuth or MnBi phase remaining in the after-measurement sample 1 of this study, which gives a higher Nernst thermopower (maximum $S_{yx} \sim 110$ μV K$^{-1}$) than that of sample 2.

The magnetic field dependence of the magnetization was extensively studied in samples heated up to 390 K, and no trace of hysteresis was observed in the paramagnetic regime (Supplementary Fig. 1f), as would be expected if the sample contains trace inclusions of MnBi, which remains ferromagnetic up to ~ 600 K.

**Magnetization behavior along the spin canting direction**

As shown in Supplementary Fig. 2, with measurement field parallel to spin canting direction [110], the temperature dependence of the magnetization mimics what is presented in ref. 1, which shows a sharp change in signals around 250 K in FC (field cooling) mode while not in ZFC (zero field cooling) mode, signaling the presence of spin canting at the same temperature as the previous report[1]; the field dependence of magnetization can be described as follows: at near-zero fields, magnetization demonstrates sharp "turn-on" features, which is likely to be the signature of the spin canting magnetic structure, and such low-field characteristics persist through a wide temperature range and finally go away when temperature rises above the spin canting temperature $T_{SC}$. Unlike ref. 1 where with a measurement field in the *ab* plane a magnetic hysteretic loop was observed but without near-zero-field features like ours, all the field dependence magnetization curves of this study pass through the origin and there is no recognizable magnetic hysteretic behavior.

It is noted that in this case we did not adopt μ$_B$/f.u. as the units for magnetization. Instead,

emu/g is used to show our considerations that the contributions to the magnetic moments should include not only the magnetic atoms but also the itinerant electrons in the system.

**<u>Electrical characterization of the samples</u>**

The low-field Hall effect ($\mu_0 H \ll 0.1$ T) was used to determine the Hall density ($-1/R_H e$) and Hall mobility ($|R_H/\rho_{xx}(H=0)|$). The results are given in Supplementary Fig. 3. In a two-carrier system, this is not a reliable assessment of the density of holes and electrons but in the low-field limit it gives a good indication of the concentration of the highest-mobility carrier. An additional complication here is the presence of a strong Berry-curvature-induced AHE, but since this is a property of the band structure, the temperature dependence of $-1/R_H e$ and $|R_H/\rho_{xx}(H=0)|$ still represent that of the highest-mobility carrier. The samples are *n*-type. The temperature dependence of $-1/R_H e$ is characteristic of an intrinsic semiconductor or semimetal with a very small or zero band overlap. The temperature dependence of $|R_H/\rho_{xx}(H=0)|$ suggests that phonons are the dominant scatterers.

To illustrate the sample-to-sample reproducibility of the results in the main figures, the same thermomagnetic properties were measured on sample 2 as what are given in the main figures on sample 1. Supplementary Figs. 4, 6, and 7 follow the same structure as Fig. 2, 3, and 4 of the main text but give the data on sample 2. The results reproduce those on sample 1 remarkably well, in spite of sample 2 having larger Hall density and lower Hall mobilities. The differences are attributed to the fact that the chemical potential is farther from the Weyl points in sample 2.

Supplementary Figure 4 shows the transverse Nernst and Hall effects on sample 2. In particular, Fig. 2a of the main text and Supplementary Fig. 4a show the Nernst thermopower of samples 1 and 2, respectively, while Fig. 2d of the main text and Supplementary Fig. 4e show the Nernst conductivity.

# Mathematical derivation of $\alpha_{yx}$ and $S_{yx}$

The Nernst thermopower and Nernst conductivity are related as follows:

From the phenomenological linear transport equations, we have:

$$\begin{pmatrix} \vec{E} \\ \vec{J_Q} \end{pmatrix} = \begin{pmatrix} \overleftrightarrow{\rho} & \overleftrightarrow{S} \\ \overleftrightarrow{\Pi} & -\overleftrightarrow{\kappa} \end{pmatrix} \begin{pmatrix} \vec{J_C} \\ \vec{\nabla T} \end{pmatrix} \tag{S1}$$

where $\overleftrightarrow{\rho}$ is the electrical resistivity tensor, $\overleftrightarrow{S}$ is the Seebeck-Nernst coefficient tensor, $\overleftrightarrow{\Pi}$ is the Peltier-Ettingshausen coefficient tensor, and $\overleftrightarrow{\kappa}$ is the thermal conductivity tensor.

Linear Onsager relations can be in the form:

$$\begin{pmatrix} \vec{J_C} \\ \vec{J_Q} \end{pmatrix} = \begin{pmatrix} \overleftrightarrow{\sigma} & -\overleftrightarrow{\alpha} \\ \overleftrightarrow{L_{TE}} & -\overleftrightarrow{\kappa} \end{pmatrix} \begin{pmatrix} \vec{E} \\ \vec{\nabla T} \end{pmatrix} \tag{S2}$$

where $\overleftrightarrow{\sigma}$ is the electrical conductivity tensor, $\overleftrightarrow{\alpha}$ is the Seebeck-Nernst conductivity tensor, and $\overleftrightarrow{L_{TE}}$ is the Peltier-Ettingshausen conductivity tensor. $\overleftrightarrow{L_{TE}}(H) = T\overleftrightarrow{\alpha}^T(-H)$.

From Eq. (S1), the electric field can be expressed as:

$$\vec{E} = \overleftrightarrow{\rho}\vec{J_C} + \overleftrightarrow{S}\vec{\nabla T} \tag{S3}$$

while the charge flux $\vec{J_C}$ can be obtained from Eq. (S2) in the form:

$$\vec{J_C} = \overleftrightarrow{\sigma}\vec{E} - \overleftrightarrow{\alpha}\vec{\nabla T} \tag{S4}$$

In the absence of a charge flux, $\vec{J_C} = 0$, Eq. (S3) and Eq. (S4) will become:

$$\vec{E} = \overleftrightarrow{S}\vec{\nabla T} \tag{S5}$$

$$\overleftrightarrow{\alpha}\vec{\nabla T} = \overleftrightarrow{\sigma}\vec{E} \tag{S6}$$

Plugging Eq. (S5) into Eq. (S6) will lead us to:

$$\overleftrightarrow{\alpha} = \overleftrightarrow{\sigma}\overleftrightarrow{S} \tag{S7}$$

Because of $\overleftrightarrow{\sigma} = \overleftrightarrow{\rho}^{-1}$, Eq. (S7) now can be written as:

$$\overleftrightarrow{\alpha} = \overleftrightarrow{\rho}^{-1}\overleftrightarrow{S} \tag{S8}$$

In our experimental configuration, applied heat or charge flux is along $x$ direction; the measured transverse voltage is along $y$ direction; the external magnetic field is along $z$ direction.

Eq. (S8) can be expanded as:

$$\begin{pmatrix} \alpha_{xx} & \alpha_{xy} \\ \alpha_{yx} & \alpha_{yy} \end{pmatrix} = \begin{pmatrix} \rho_{xx} & \rho_{xy} \\ \rho_{yx} & \rho_{yy} \end{pmatrix}^{-1} \begin{pmatrix} S_{xx} & S_{xy} \\ S_{yx} & S_{yy} \end{pmatrix} = \frac{1}{\rho_{xx}\rho_{yy} - \rho_{xy}\rho_{yx}} \begin{pmatrix} \rho_{yy} & -\rho_{xy} \\ -\rho_{yx} & \rho_{xx} \end{pmatrix} \begin{pmatrix} S_{xx} & S_{xy} \\ S_{yx} & S_{yy} \end{pmatrix}$$

$$\begin{pmatrix} \alpha_{xx} & \alpha_{xy} \\ \alpha_{yx} & \alpha_{yy} \end{pmatrix} = \begin{pmatrix} \dfrac{\rho_{yy}S_{xx} - \rho_{xy}S_{yx}}{\rho_{xx}\rho_{yy} - \rho_{xy}\rho_{yx}} & \dfrac{\rho_{yy}S_{xy} - \rho_{xy}S_{yy}}{\rho_{xx}\rho_{yy} - \rho_{xy}\rho_{yx}} \\ \dfrac{-\rho_{yx}S_{xx} + \rho_{xx}S_{yx}}{\rho_{xx}\rho_{yy} - \rho_{xy}\rho_{yx}} & \dfrac{-\rho_{yx}S_{xy} + \rho_{xx}S_{yy}}{\rho_{xx}\rho_{yy} - \rho_{xy}\rho_{yx}} \end{pmatrix} \quad (S9)$$

On the basis of the measured elements in $\overleftrightarrow{\rho}$ and $\overleftrightarrow{S}$, the Nernst conductivity can be represented in the formula:

$$\alpha_{yx} = \frac{-\rho_{yx}S_{xx} + \rho_{xx}S_{yx}}{\rho_{xx}\rho_{yy} - \rho_{xy}\rho_{yx}} \quad (S10)$$

where $S_{xx}$ and $S_{yx}$ are the Seebeck coefficient and Nernst coefficient, respectively; $\rho_{xx}$, $\rho_{yx}$, and $\rho_{yy}$ are electrical resistivity along $x$ direction, electrical Hall resistivity, and electrical resistivity along $y$ direction, respectively.

According to Akgoz-Saunders' theory[2], which combines crystal symmetry with Onsager's reciprocity, in YbMnBi$_2$ and our experimental configurations, $\rho_{xy}(H) = -\rho_{yx}(H)$ is expected. Thus, the determinant of the electrical resistivity matrix can be reduced as: $\rho_{xx}\rho_{yy} - \rho_{xy}\rho_{yx} = \rho_{xx}\rho_{yy} + \rho_{yx}^2$, which leads to a simplified form of Eq. (S10):

$$\alpha_{yx} = \frac{-\rho_{yx}S_{xx} + \rho_{xx}S_{yx}}{\rho_{xx}\rho_{yy} + \rho_{yx}^2} \quad (S11)$$

Rewriting Eq. (S10):

$$\alpha_{yx} = \frac{-\rho_{yx}}{\rho_{xx}\rho_{yy} - \rho_{xy}\rho_{yx}} S_{xx} + \frac{\rho_{xx}}{\rho_{xx}\rho_{yy} - \rho_{xy}\rho_{yx}} S_{yx} = \sigma_{yx} S_{xx} + \sigma_{yy} S_{yx}$$

$$S_{yx} = \frac{\alpha_{yx} - \sigma_{yx} S_{xx}}{\sigma_{yy}} \quad (S12)$$

**Adiabatic versus isothermal Seebeck and Nernst coefficients**

Based on the phenomenological linear transport equations, when there is no charge flux applied, under the experimental configuration of this work, Eq. (S5) can be unfolded as:

$$\begin{pmatrix} E_x \\ E_y \end{pmatrix} = \begin{pmatrix} S_{xx} & S_{xy} \\ S_{yx} & S_{yy} \end{pmatrix} \begin{pmatrix} \nabla_x T \\ \nabla_y T \end{pmatrix} \quad (S13)$$

All the components in the above Seebeck-Nernst coefficient tensor are isothermal thermoelectric power.

In the adiabatic measurement conditions, starting from the definitions of Seebeck and Nernst coefficients, we have:

$$S_{xx}^{Adiabatic} = \frac{E_x}{\nabla_x T}\Big|_{\overrightarrow{J_{cx}},\overrightarrow{J_{cy}},\overrightarrow{J_{Qy}}=0} = \frac{S_{xx}\nabla_x T + S_{xy}\nabla_y T}{\nabla_x T} = S_{xx} + S_{xy}\frac{\nabla_y T}{\nabla_x T} \quad (S14)$$

$$S_{yx}^{Adiabatic} = \frac{E_y}{\nabla_x T}\Big|_{\overrightarrow{J_{cx}},\overrightarrow{J_{cy}},\overrightarrow{J_{Qy}}=0} = \frac{S_{yx}\nabla_x T + S_{yy}\nabla_y T}{\nabla_x T} = S_{yx} + S_{yy}\frac{\nabla_y T}{\nabla_x T} \quad (S15)$$

Clearly, $S_{xx}^{Adiabatic}$ equals its isothermal counterpart $S_{xx}$ plus a contribution related to a Nernst thermopower $S_{xy}$ and the tangent of thermal Hall angle $tan\theta \equiv \nabla_y T/\nabla_x T$. Similarly, $S_{yx}^{Adiabatic}$ equals its isothermal counterpart $S_{yx}$ plus a contribution related to a Seebeck thermopower $S_{yy}$ and the tangent of thermal Hall angle $tan\theta \equiv \nabla_y T/\nabla_x T$.

In this research, $|\nabla_y T/\nabla_x T| < 0.06$ (see Fig. 4c of the main text and Supplementary Fig. 7c), and according to ref. 1, $S_{xy}$ and $S_{yy}$ are in similar or smaller magnitudes compared with the adiabatic coefficients on the left of Eq. (S14) and Eq. (S15), respectively. Therefore, $S_{xx}^{Adiabatic} \approx S_{xx}$ (isothermal) and $S_{yx}^{Adiabatic} \approx S_{yx}$ (isothermal), in the context of this study.

**Nernst conductivity for a partially filled Weyl band**
The Nernst conductivity $\alpha_{yx}$ is generically related to the Hall conductivity $\sigma_{yx}$ by

$$\alpha_{yx} = \frac{1}{eT}\int dE(E-\mu)\left(-\frac{df}{dE}\right)\sigma_{yx}(E) \quad (S16)$$

where $\mu$ is the chemical potential (which we define relative to the Weyl point), $f(E,\mu)$ is the Fermi function, and $\sigma_{yx}(E)$ denotes the Hall conductivity in the limit of zero temperature when the chemical potential is equal to $E$ (ref. 3). In the limit where $k_B T$ is much smaller than $|\mu|$, a Sommerfeld expansion gives

$$\alpha_{yx} \simeq \frac{\pi^2 k_B^2 T}{3 e}\left(\frac{d\sigma_{yx}}{dE}\right)_{E=\mu} \quad (S17)$$

Thus the Nernst conductivity at low temperature reflects the derivative of the Hall conductivity $\sigma_{yx}$ with respect to the chemical potential $\mu$, and any contribution to $\sigma_{yx}$ that is independent of $\mu$ does not affect $\alpha_{yx}$. Specifically, at low temperature the anomalous Hall conductivity arising from filled Weyl bands below the chemical potential does not increase the value of $\alpha_{yx}$. Consequently it is possible for the Nernst coefficient, Eq. (S12), to remain large in magnitude due to a large anomalous contribution to $\sigma_{yx}$ even while $\alpha_{yx}$ is small and has no significant contribution from filled bands or from Berry curvature.

Following the derivation in (ref. 4), we can write the Hall conductivity of a partially filled Weyl band as

$$\sigma_{yx}(E) = \frac{1}{3} \frac{e^2 N(E) v_y v_x \tau}{1 + \omega_c(E)^2 \tau^2} \omega_c(E) \tau + \sigma^{AHE} \Theta(E) \tag{S18}$$

where $\sigma^{AHE}$ is the AHE contribution to $\sigma_{yx}$ arising from filled Weyl bands, $\Theta(E)$ is the Heaviside step function,

$$N(E) = \frac{gE^2}{2\pi^2 \hbar^3 v_x v_y v_z} \tag{S19}$$

is the density of states per unit volume at energy $E$, with $g$ the degeneracy of the Weyl/Dirac point ($g = 8$ for YbMnBi2), $\tau$ is the momentum relaxation time, $\hbar$ is the reduced Planck constant,

$$\omega_c(E) = \frac{eB v_x v_y}{E} \tag{S20}$$

is the cyclotron frequency for motion in the *x-y* plane (which, for a linear dispersion relation, depends on the electron energy $E$), and $v_x$, $v_y$, $v_z$ are the Weyl velocities along the *x, y,* and *z* directions, respectively. Using Eq. (S17) at $\mu \neq 0$ gives the following formula for the low-temperature Nernst conductivity:

$$\alpha_{yx} \simeq \frac{g}{18} \frac{e^2 k_B^2 T v_x v_y \tau^2 B}{v_z \hbar^3} \frac{1 + 3\omega_c^2 \tau^2}{(1 + \omega_c^2 \tau^2)^2} \tag{S21}$$

which generalizes the result of Eq. (39) in ref. 4.

In terms of the dependence on magnetic field $B$, Eq. (S21) can be written as

$$\alpha_{yx} \simeq \frac{A_0 T}{B_0^2} B \frac{1 + 3(B/B_0)^2}{(1 + (B/B_0)^2)^2} \tag{S22}$$

where $A_0$ and $B_0$ are constants given by

$$A_0 = \frac{g k_B^2 \mu^2}{18 \hbar^3 v_x v_y v_z} \tag{S23}$$

and

$$B_0 = \frac{|\mu|}{e v_x v_y \tau}. \tag{S24}$$

Fig. 2e of the main text and Supplementary Fig. 4f show the theoretical result of Eq. (S22) for $\alpha_{yx}$ as compared to the experimental result, with $A_0$ and $B_0$ treated as fitting parameters for each temperature. The fitting provides a good quantitative description of the measurements in both sample 1 (Fig. 2d of the main text) and sample 2 (Supplementary Fig. 4e). The fitted values of $A_0$ and $B_0$ are given for each sample and each temperature in Supplementary Table 2.

The values of the fitted parameters $A_0$ and $B_0$ imply estimates for the chemical potential $\mu$ and scattering time $\tau$, provided that one has an estimate for the band velocities $v_x$, $v_y$, $v_z$. Previous studies of YbMnBi$_2$ using angle-resolved photoemission spectroscopy[5] and optical conductivity[6] arrived at estimates for the Fermi velocities $\hbar v_x \approx \hbar v_z \approx 9$ eVÅ and $\hbar v_y \simeq 0.043$ eVÅ, or equivalently $v_x \approx v_z \approx 1.4 \times 10^6$ m/s and $v_y \approx 6.5 \times 10^3$ m/s. This huge anisotropy in velocities is consistent with our observed anisotropy of the resistivity, which suggests that these highly anisotropic Weyl pockets are indeed primarily responsible for the transport at low temperature. Theoretically, at low temperature $\rho_{yy}/\rho_{xx} = v_x^2/v_y^2$, or $v_x/v_y = \sqrt{\rho_{yy}/\rho_{xx}}$. Our measured resistivities at $T = 43$ K for sample 1 are $\rho_{yy} \approx 5.8 \times 10^{-4}$ $\Omega$ m and $\rho_{xx} \approx 1.1 \times 10^{-7}$ $\Omega$ m (see Fig. 3 of the main text), which gives an estimate $v_x/v_y \approx 70$, as compared to $v_x/v_y \approx 200$ for the studies cited above. In order to reduce the number of fitting parameters in our estimates, we use the literature values of $v_x$, $v_y$, and $v_z$ in the following estimates for $\mu$ and $\tau$.

Focusing on the lowest temperature of our thermoelectric measurements, $T = 43$ K, Eq. (S23) implies $\mu \approx 1.3$ meV for sample 1 and $\mu \approx 10.2$ meV for sample 2, consistent with our interpretation that sample 1 has a chemical potential that is considerably closer to the Weyl points. From Eq. (S24) we can estimate the corresponding momentum relaxation times as $\tau \approx 0.31$ ps for sample 1 and $\tau \approx 0.20$ ps for sample 2, which is again consistent with our interpretation that sample 2 has a higher concentration of unintentional dopant impurities and therefore a shorter momentum relaxation time[7].

We can now consider the temperature dependence of the parameters $A_0$ and $B_0$. In YbMnBi$_2$ the light, high-mobility Weyl bands coincide with heavy, low-mobility hole bands. Since the heavy hole bands have a much higher density of states, as the temperature increases the chemical potential moves upward toward the top of the hole band in order to maintain the total charge density (holes minus electrons) in the system. Consequently, the value of $\mu$ increases with temperature. Increased temperature also increases the rate of electron-phonon scattering, so that $\tau$ declines. Consequently, we expect the values of $A_0$ and $B_0$ to increase with temperature. This is consistent with what is seen for sample 1, for which one can infer that the value of $\mu$ increases by about 2.5 times from $T = 43$ K to $T = 304$ K while the value of $\tau$ is reduced by about 5 times. On the other hand, sample 2 shows a different dependence, with $A_0$ *decreasing* with temperature and $B_0$ having a weak temperature dependence (first decreasing slightly and then increasing again).

One possible explanation for this dependence is that for the larger chemical potential associated with sample 2 the dispersion relation is no longer strictly linear in momentum, so that the Fermi velocities $v_x$, $v_y$, $v_z$ have a significant dependence on the value of $\mu$. Such additional, parabolic terms in the dispersion can be expected to cause the Fermi velocities to increase as the chemical potential rises with temperature, which would cause the values of both $A_0$ and $B_0$ to decline. We suspect that in sample 2 we are seeing this effect overwhelming the increase in $\mu$ in the $T$-dependence of $A_0$, while for the coefficient $B_0$ we observe a competition between the increase in the band velocities and the decrease in $\tau$ with increasing $T$.

**<u>Interpretation of the Hall resistivity</u>**

In a common conducting system with a single trivial band of $n$-type carriers, the Hall resistivity $\rho_{xy} = -\rho_{yx} = B/ne$, where $B$ is the magnetic field and $n$ is the carrier concentration. When there are coexisting electron and hold bands, as in the case of YbMnBi$_2$, the general expression for the Hall resistivity is

$$\rho_{xy} = \frac{B}{e} \frac{n_h \mu_h^2 - n_e \mu_e^2 + B^2 \mu_e^2 \mu_h^2 (n_h - n_e)}{n_e^2 \mu_e^2 (1 + B^2 \mu_h^2) + n_h^2 \mu_h^2 (1 + B^2 \mu_e^2) + 2 n_e n_h \mu_e \mu_h (1 - B^2 \mu_e \mu_h)} \qquad (S25)$$

where $n_e$, $n_h$ and $\mu_e$, $\mu_h$ are the carrier concentrations and mobilities of the electron and hole bands, respectively[8]. For a system with a small concentration of high-mobility electrons and a large concentration of low-mobility holes, as in YbMnBi$_2$, Eq. (S25) describes a Hall resistivity with a non-monotonic slope as a function of $B$. At small fields, the Hall effect is $n$-type provided that $n_e \mu_e^2 \gg n_h \mu_h^2$, while at large enough fields that $B^2 \mu_e^2 \gg 1$ the magnitude of $\rho_{xy}$ reaches a peak and declines again. This behavior can be seen in Fig. 2b of the main text. The sign of the derivative $d\rho_{yx}/dB$ at high fields reflects the dominant carrier type. At low temperatures, the very high mobility of the Weyl bands allows them to remain dominant at high fields, so that the slope of $\rho_{yx}$ is negative ($n$-type). At sufficiently high temperatures that $k_B T$ is greater than or comparable to the chemical potential of the Weyl bands, however, thermally excited holes proliferate in the Weyl valence band, providing a high-mobility channel for *p*-type conduction as well. Thus, the combination of high-mobility holes in the Weyl bands and a large concentration of low-mobility holes in the trivial bands allows the *p*-type channel to overcome the *n*-type conduction channel and the sign of the Hall effect is reversed. This effect is apparently seen in Fig. 2b. Notice that in sample 2 the sign of $d\rho_{yx}/dB$ is never completely reversed at high temperature (see Supplementary Fig. 4c), which is consistent with our interpretation that the chemical potential is farther away from the Weyl points and therefore holes can be thermally excited in the Weyl bands only at significantly higher temperatures.

In the limit of low field, Taylor expanding Eq. (S25) gives

$$\rho_{xy} = -\frac{B}{e} \frac{\mu_e^2 n_e + \mu_h^2 n_h}{(\mu_e n_e + \mu_h n_h)^2} \tag{S26}$$

so that the Hall concentration of electrons becomes

$$-\frac{1}{R_H e} = -\frac{B}{e\rho_{xy}} = \frac{(\mu_e n_e + \mu_h n_h)^2}{\mu_e^2 n_e + \mu_h^2 n_h} \simeq n_e + \frac{2 n_h \mu_h}{\mu_e} \tag{S27}$$

where the last equality in Eq. (S27) corresponds to the limit $\mu_e n_e \gg \mu_h n_h$ (transport dominated by *n*-type carriers). As the temperature increases, the quantity -1/($R_H$ *e*) evolves due to the changing mobilities of the two bands and the shifting position of the chemical potential. On the one hand, increasing temperature causes the mobility of the hole band to decline quickly, since phonon scattering is more effective in the heavy hole band due to its larger density of states, and this declining mobility $\mu_h$ appears as a decrease in -1/($R_H$ *e*). On the other hand, increasing temperature causes the chemical potential to rise toward the top of the hole band and away from the Weyl points. Specifically, when a light electron band and heavy hole band simultaneously intersect the chemical potential, the chemical potential evolves with increased temperature in such a way as to maintain electroneutrality (i.e., to maintain a constant difference between the electron and hole concentrations). Since in our case the trivial band of heavy holes has a significantly higher density of states than the light Weyl band, the chemical potential moves upward toward the top of the hole band as temperature increases. This movement produces an effective increase in the concentration of electrons in the Weyl band, leading to an apparent increase in $-1/(R_H e)$.

We interpret the nonmonotonic variation of the Hall density with temperature, as shown in Supplementary Fig. 3, as a competition between these two effects.

**The ratio of $|\alpha_{ij}/\sigma_{ij}|$**

As presented in Supplementary Fig. 8, the ratio of the total Nernst conductivity $\alpha_{yx}$ to the total Hall conductivity $\sigma_{yx}$ of the present two samples shows consistency that for temperatures below 80 K throughout the whole experimental field range the ratio is basically no greater than the unity of $k_B/e$ while for the other measured temperatures there are a considerable portion of the data, even not all of them, which are no less than the unity of $k_B/e$ across the experimental field points.

When drawing comparisons with other compounds in terms of the temperature dependence of the ratio $|\alpha_{ij}/\sigma_{ij}|$, the values of the present samples were selected at magnetic fields where $\alpha_{yx}$ reaches maximum, which is the protocol argued by Tang et al. when they did the same thing to SmMnBi$_2$ (ref. 9). As proposed by Behnia et al.[10], the ratio of the anomalous Nernst conductivity to the anomalous Hall conductivity approaches the unity of $k_B/e$ when $T \sim 300$ K if the dominant contributions to both properties are in the origin of Berry curvature. It is found that the ratio for the present two samples is within a few $k_B/e$ although it increases from less than the unity of $k_B/e$ to several multiples of $k_B/e$ as temperature rises. This is in direct contrast with the previous study (ref. 1), in which the corresponding ratio is a few hundreds of $k_B/e$ in the experimental configuration where their ANE conductivity $|\alpha_{ANE}| \sim 10$ A m$^{-1}$ K$^{-1}$ was claimed to be the highest

amongst all AFM materials. The observation here suggests that the anomalous transverse transport responses in this study are of much more intrinsic origin.

**Thermoelectric figure of merit $z_{xx}T$ and $z_{yx}T$**

By definition, the longitudinal thermoelectric figure of merit is $z_{xx}T = \frac{S_{xx}^2 T}{\kappa_{xx}\rho_{xx}}$, while the transverse thermoelectric figure of merit is $z_{yx}T = \frac{S_{yx}^2 T}{\kappa_{xx}\rho_{yy}}$. As shown in Supplementary Fig. 9, the magnitudes of $z_{xx}T$ and $z_{yx}T$ of the present two samples are small. Although the Nernst thermopower $S_{yx}$ in this study is large, the heavy effective mass of charge carriers along the $y$ direction leads to enormous $\rho_{yy}$, which puts out the hope for a considerable transverse thermoelectric figure of merit. It is possible that the alternative transverse figure of merit $z_{xy}T = \frac{S_{xy}^2 T}{\kappa_{yy}\rho_{xx}}$ is significantly larger, due to the much smaller resistance $\rho_{xx}$ along the $x$ direction, but unfortunately our current measurement setup and samples' geometry do not enable us to directly measure the quantities $S_{xy}$ or $\kappa_{yy}$.

**Transport properties of sample 1 after 60-degree rotation against field**

Given that the Nernst response we observe is much larger than the one observed by Pan et al.[1], it is natural to ask whether the alignment of the magnetic field with the spin canting direction, [110], is crucial for achieving a large Nernst signal. To address this question, we perform an experiment in which the sample is rotated with respect to the field direction, so that the magnetic field remains in the *ab* plane but is no longer aligned with the [110] direction. As displayed in Supplementary Fig. 10, the field dependence of the Seebeck coefficient $S_{xx}$, thermal resistivity $w_{xx}$, and thermal conductivity $\kappa_{xx}$ at various temperatures is quite similar to the case where the field is aligned along [110]. The value of the Nernst thermopower $S_{yx}$ shows an appreciable drop in magnitude of $S_{yx}$, which is consistent with the fact that the component of the external field along the direction [110] is reduced. However, even so, at $T > 250$ K, the high-field saturation values $\geq 25$ µV K$^{-1}$ of $S_{yx}$ are still quite large, especially when compared to those of the compounds listed in Table 1. The curve $S_{yx}(B)$ is remarkably similar in shape for both the rotated (Supplementary Fig. 10b) and unrotated (Fig. 2a of the main text) case. These observations allow us to confidently reject the possibility that the giant anomalous Nernst thermopower we observed is due solely to the alignment of the external field with the spin canting direction [110], and confirm our understanding that the much lower chemical potential $\mu$ in the present samples plays the key role.

**Temperature dependence of $S_{xx}$ and $\rho_{xx}$**

Supplementary Fig. 11 displays the temperature dependence of the Seebeck coefficient $S_{xx}$ and electrical resistivity $\rho_{xx}$ of the two samples in the present study. Drawing contrast with the previous report[1], the observations here are not only qualitatively but also quantitatively comparable.

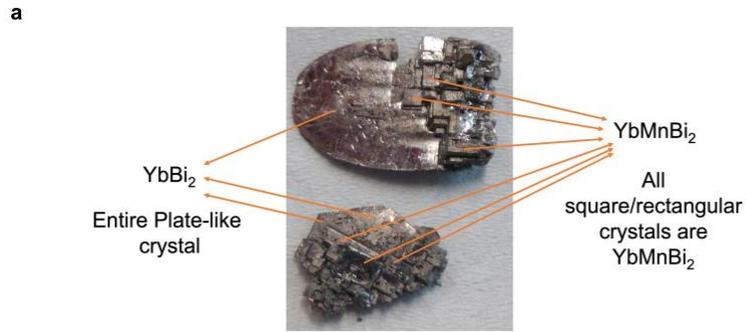

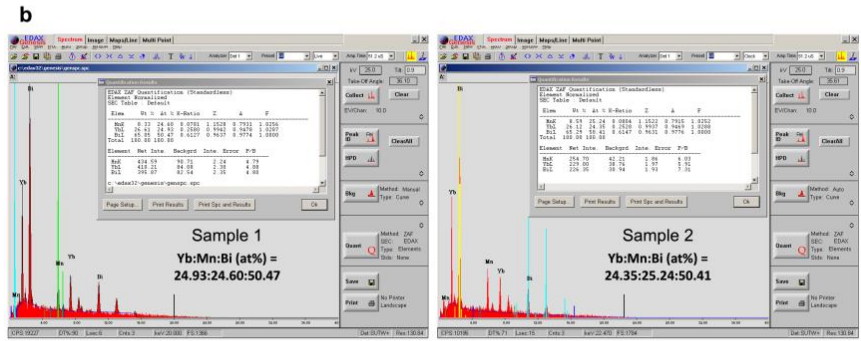

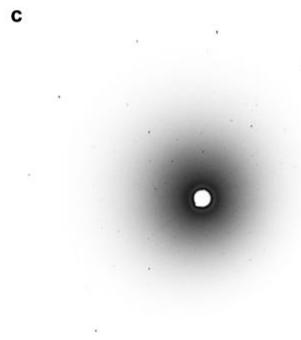
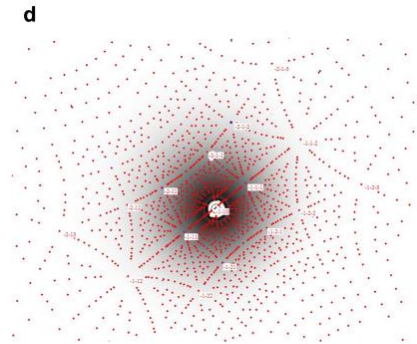

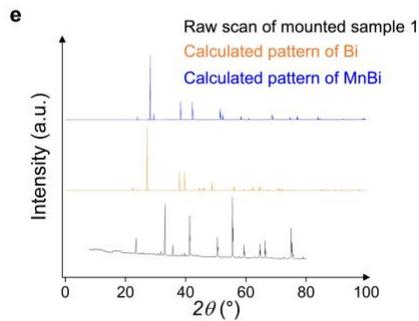
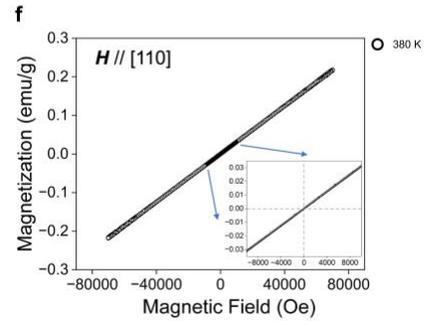

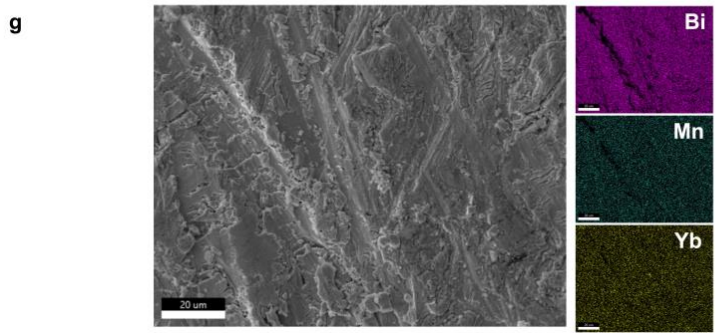

**Supplementary Fig. 1 | Optical image, composition, structure, and phase characterization as well as magnetic hysteresis detection. a**, Optical image of the uncut unseparated as-grown YbMnBi$_2$ crystal cluster. **b**, Energy-dispersive X-ray analysis (EDAX) demonstrates the compositional atomic ratios of the two as-grown crystals, which aligns well with the stoichiometric ratio of 1:1:2. EDAX results for two additional single crystals are shown in Supplementary Fig. 12. **c**–**d**, Unfitted (**c**) and fitted (**d**) Laue diffraction patterns confirm the single-crystallinity of the as-grown YbMnBi$_2$ crystal. **e**, Clearly, the raw XRD scan of the after-measurement mounted sample 1 in this study (black curve) does not contain peaks from the calculated powder XRD pattern of elemental bismuth (orange curve) or MnBi compound (blue curve), indicating it is unlikely that there are precipitates of elemental bismuth or MnBi inclusions in the after-measurement sample. **f**, Magnetic field-dependence of magnetization when $H$ // [110] at 380 K. Inset shows a magnified plot of the low-field region, in which dashed reference lines crossing at the origin are added to guide the eyes. At such a temperature way above $T_N$ of YbMnBi$_2$, scanning field back and forth between -70k Oe and +70k Oe, there is no recognizable magnetic hysteresis loop, which is indicative of a remote chance of MnBi inclusions being existent in the sample. **g**, Backscattered electrons (BSE) image and elemental mapping of Bi, Mn, and Yb, which confirms the homogeneity.

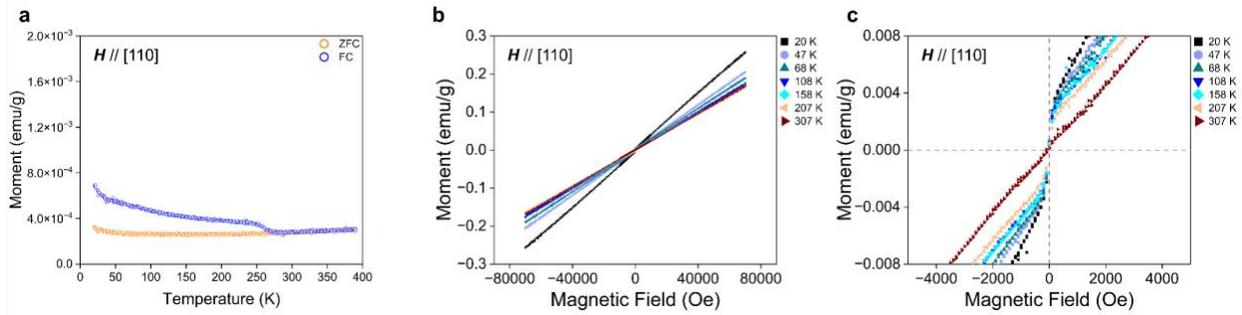

**Supplementary Fig. 2 | Temperature and field dependence of magnetization when the measurement field *H* // [110]**. **a**. Temperature dependence of magnetization under ZFC (zero field cooling) and FC (field cooling) conditions. The FC cooling field is 70 kOe and the measurement field for both ZFC and FC modes is 100 Oe. A sharp change in magnetic moment signals was observed around the reported spin canting temperature ($T_{SC}$) 250 K in FC mode while not in ZFC mode. **b**–**c**, Field dependence of magnetization at various temperatures. (**c**) is the magnified low-field region plot of (**b**). Dashed reference lines crossing at the origin were added in (**c**) as a guide to the eye. At low fields, the magnetic moment demonstrates a noticeable slope-change behavior near 0 Oe when the temperature is below $T_{SC}$, while in other regions the magnetic moment changes linearly with field. Such a distinction between field ranges was not seen for temperatures above $T_{SC}$. Across the measured temperatures, all curves pass through the origin and there is no observable magnetic hysteretic behavior.

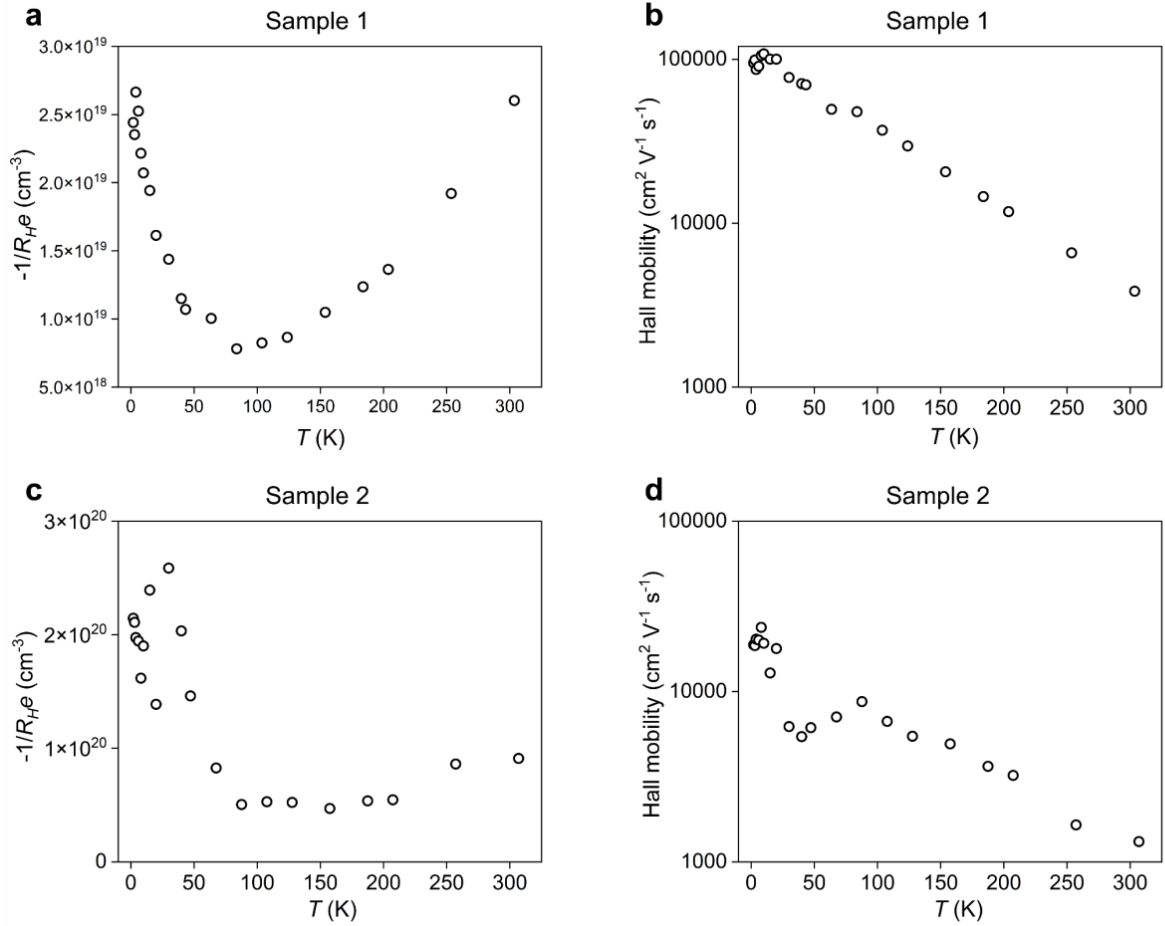

**Supplementary Fig. 3 | Electrical characterization of samples 1 and 2. a–d**, The low-field Hall density $-1/R_H e$ and Hall mobility $|R_H/\rho_{xx}(H=0)|$ of samples 1 (**a**, **b**) and 2 (**c**, **d**), respectively.

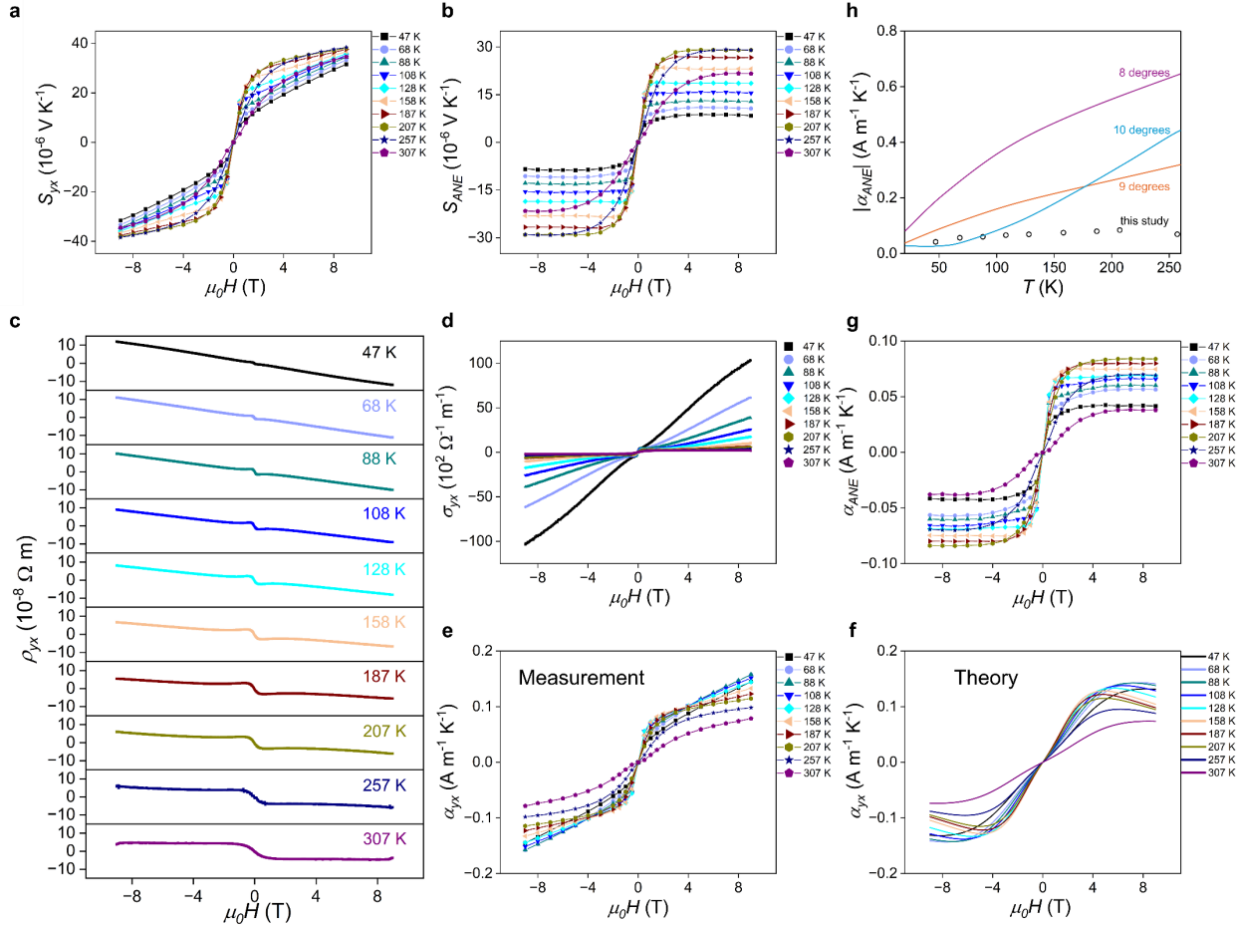

**Supplementary Fig. 4 | Transverse transport properties of sample 2. a**, Total Nernst thermopower $S_{yx}$ as a function of field at various temperatures. **b**, The anomalous contribution $S_{ANE}$ of the total Nernst coefficient was extracted following the method outlined by ref. 17. Field dependence of $S_{ANE}$ shows saturation at high fields. The maximum value of ~ 30 µV K$^{-1}$ is reached between 200 K and 260 K when $S_{ANE}$ saturates. **c**, The change of Hall resistivity $\rho_{yx}$ with magnetic field at different temperatures. The value of d$\rho_{yx}$/d$H$ of sample 1 (Fig. 2b) at high fields changes sign as temperature increases; this behavior is not observed in sample 2. **d–g**, Field dependence of (**d**) Hall conductivity $\sigma_{yx}$, (**e**) total Nernst conductivity $\alpha_{yx}$ derived from measured properties described by Eq. (1), (**f**) theory-predicted values of Nernst conductivity $\alpha_{yx}$, and (**g**) the anomalous contribution $\alpha_{ANE}$, which was extracted following the method outlined by ref. 17, of the total Nernst conductivity, at various temperatures. **h**, Temperature dependence of the maximum absolute values of the anomalous Nernst conductivity $\alpha_{ANE}$ from this study in comparison with the first-principles calculations of $\alpha_{ANE}$ from ref. 17. The numerical calculations correspond to Fermi energy $E_F = 0$ eV and are plotted for various spin canting angles.

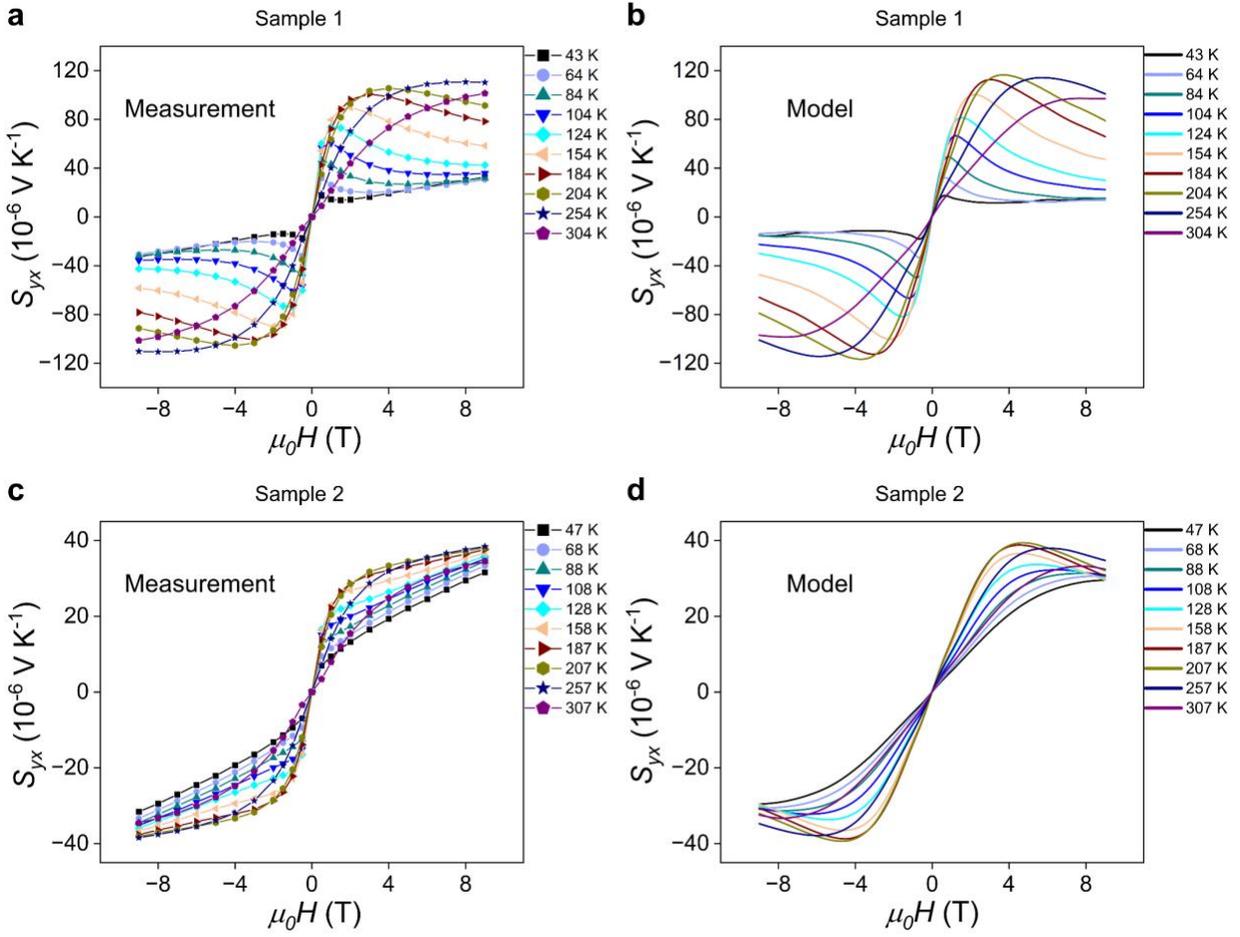

**Supplementary Fig. 5 | Comparison of measured and model-deduced Nernst thermopower $S_{yx}$ of samples 1 and 2.** Invoking Eq. (2) $S_{yx} = \frac{\alpha_{yx} - \sigma_{yx} S_{xx}}{\sigma_{yy}}$, applying the theoretical $\alpha_{yx}$ values drawn in Fig. 2e and Supplementary Fig. 4f and the measured $\sigma_{yx}$, $S_{xx}$, and $\sigma_{yy}$ values allows for the reconstruction of the Nernst thermopower $S_{yx}$. In both two samples, this semi-theoretical model is in good agreement with the measurements.

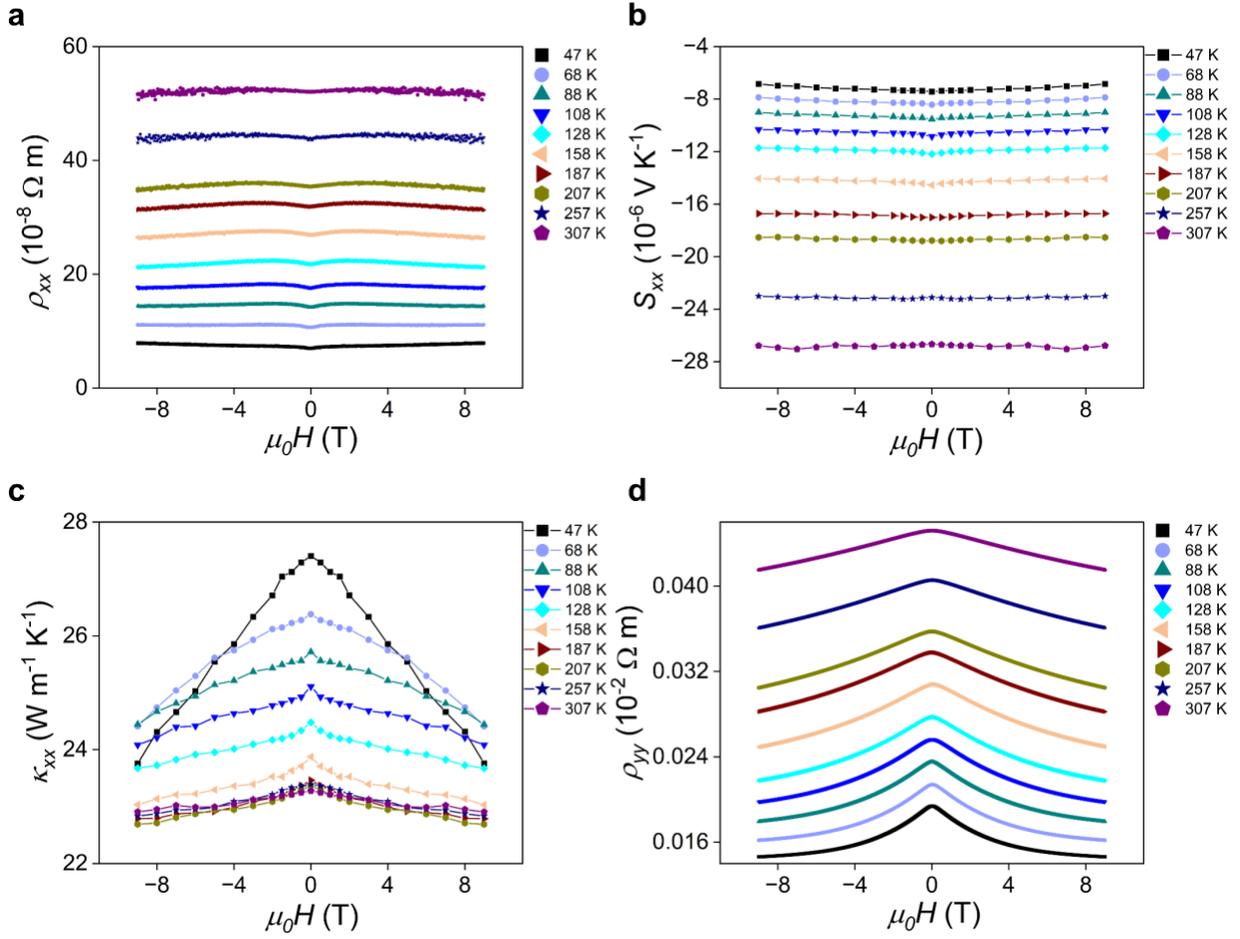

**Supplementary Fig. 6 | Longitudinal transport properties of sample 2. a–d**, Field dependence of (**a**) electrical resistivity $\rho_{xx}$, (**b**) Seebeck coefficient $S_{xx}$, (**c**) thermal conductivity $\kappa_{xx}$, and (**d**) electrical resistivity $\rho_{yy}$, at various temperatures. A temperature-robust negative magnetoresistance of up to ~ 25% at low temperatures is observed in $\rho_{yy}$.

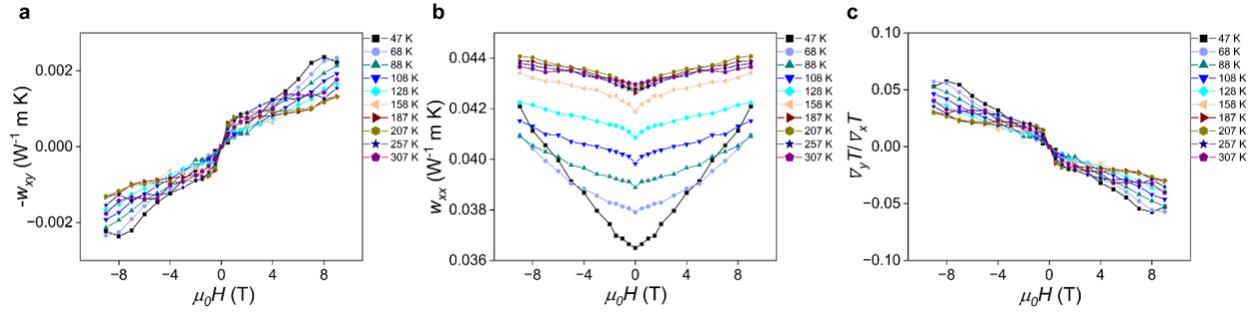

**Supplementary Fig. 7 | Appreciable thermal Hall effect signal (sample 2). a–c**, Field dependence of (**a**) thermal Hall resistivity $w_{xy}$, (**b**) thermal resistivity $w_{xx}$, and (**c**) the tangent of thermal Hall angle $\nabla_y T/\nabla_x T$, at various temperatures. The signal reproduces the results on sample 1 (Fig. 4).

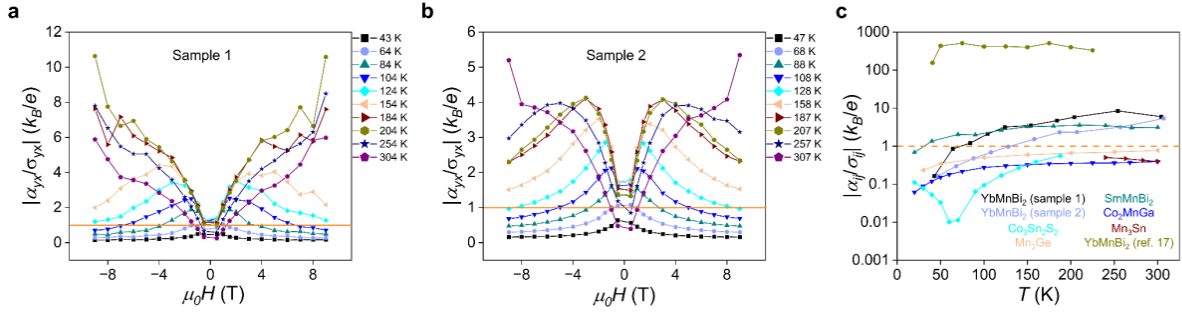

**Supplementary Fig. 8 | Field dependence of the ratio $|\alpha_{yx}/\sigma_{yx}|$ of the present two samples at various temperatures and temperature-dependence comparison with other compounds. a–b**, field dependence of the ratio $|\alpha_{yx}/\sigma_{yx}|$ at different temperatures for (**a**) sample 1 and (**b**) sample 2. The orange reference line at the unity of $k_B/e$ was added as a guide to the eye. **c**, as suggested by Tang et al.[40], we plot the ratio $|\alpha_{ij}/\sigma_{ij}|$ against $T$ for the present two samples, with each data point taken at the value of magnetic field for which $\alpha_{yx}$ achieves its maximum at the given $T$. For comparison we also plot the corresponding curve for SmMnBi$_2$[40], while for the other five listed compounds in the plot (Co$_2$MnGa[41], Co$_3$Sn$_2$S$_2$[11], Mn$_3$Ge[12], Mn$_3$Sn[14], YbMnBi$_2$[17]) we plot the ratio between just the anomalous contributions, $|\alpha_{ij}^A/\sigma_{ij}^A|$. The orange reference line at the unity of $k_B/e$ was added as a guide to the eye.

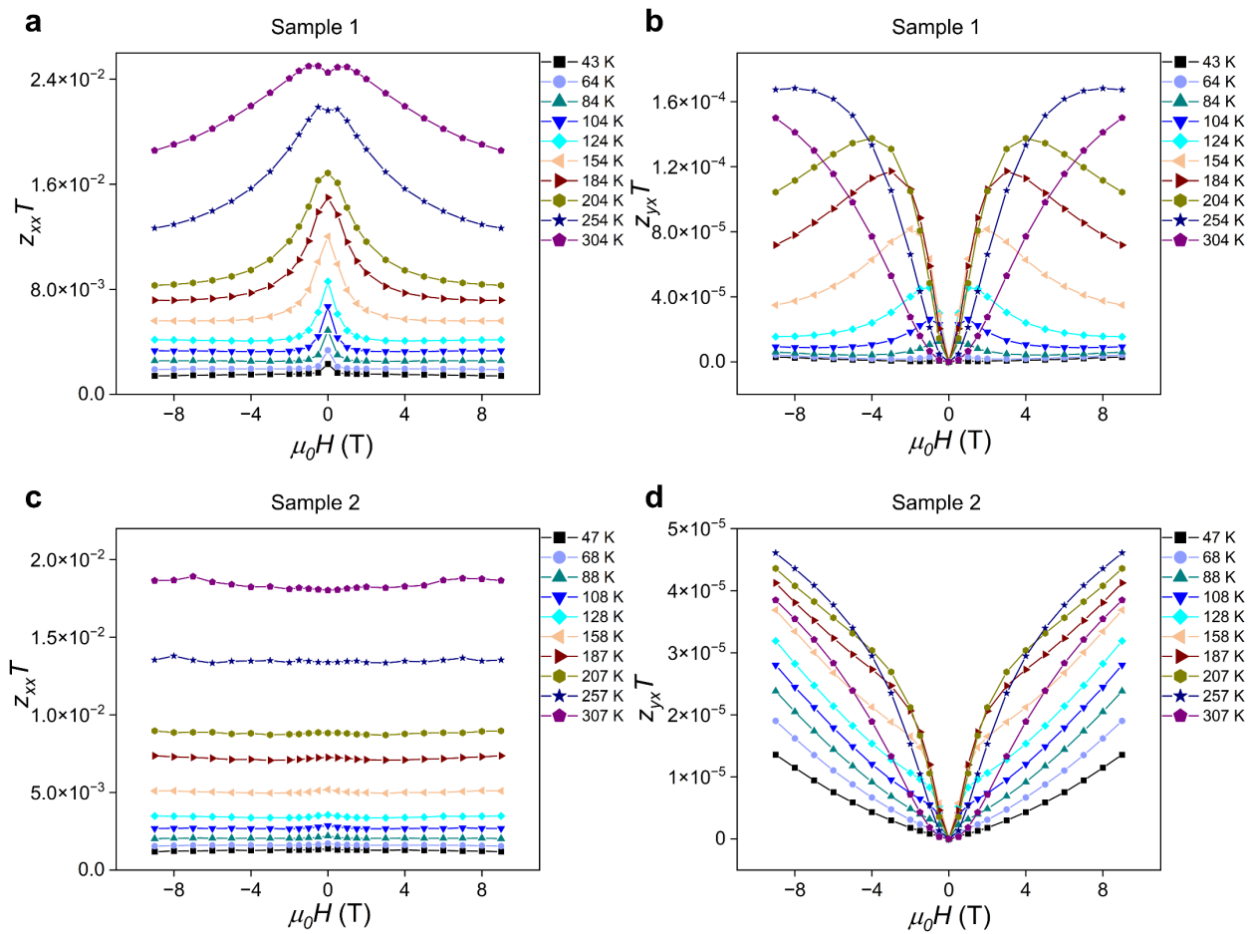

**Supplementary Fig. 9 | Field dependence of the thermoelectric figures of merit $z_{xx}T$ and $z_{yx}T$ at various temperatures for samples 1 and 2**. **a–b**, (**a**) Longitudinal thermoelectric figure of merit $z_{xx}T$ and (**b**) transverse thermoelectric figure of merit $z_{yx}T$ as a function of magnetic field at different temperatures for sample 1. **c–d**, (**c**) Longitudinal thermoelectric figure of merit $z_{xx}T$ and (**d**) transverse thermoelectric figure of merit $z_{yx}T$ as a function of magnetic field at different temperatures for sample 2.

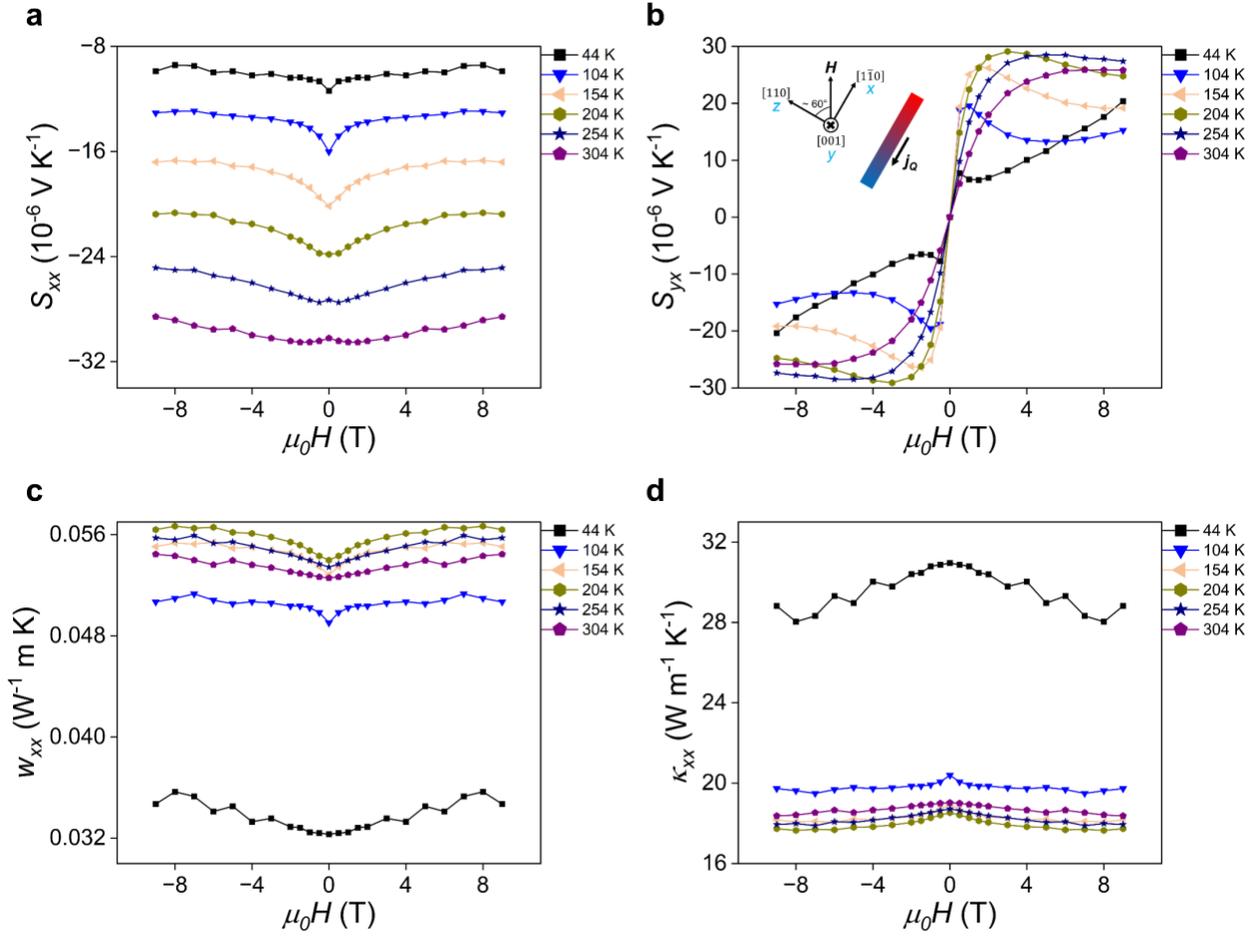

**Supplementary Fig. 10 | Transport properties of sample 1 after being rotated 60 degrees relative to the field direction**. **a**–**d**, Field dependence of (**a**) Seebeck coefficient $S_{xx}$, (**b**) Nernst coefficient $S_{yx}$, (**c**) thermal resistivity $w_{xx}$, and (**d**) thermal conductivity $\kappa_{xx}$ at various temperatures. The inset of (**b**) shows the experimental configuration, in which the external magnetic field remains in the *ab* plane but now deviates from the [110] direction by ~ 60 degrees. Heat flux $j_Q$ was applied still along [1$\bar{1}$0] and transverse voltage *V* was recorded still along [001].

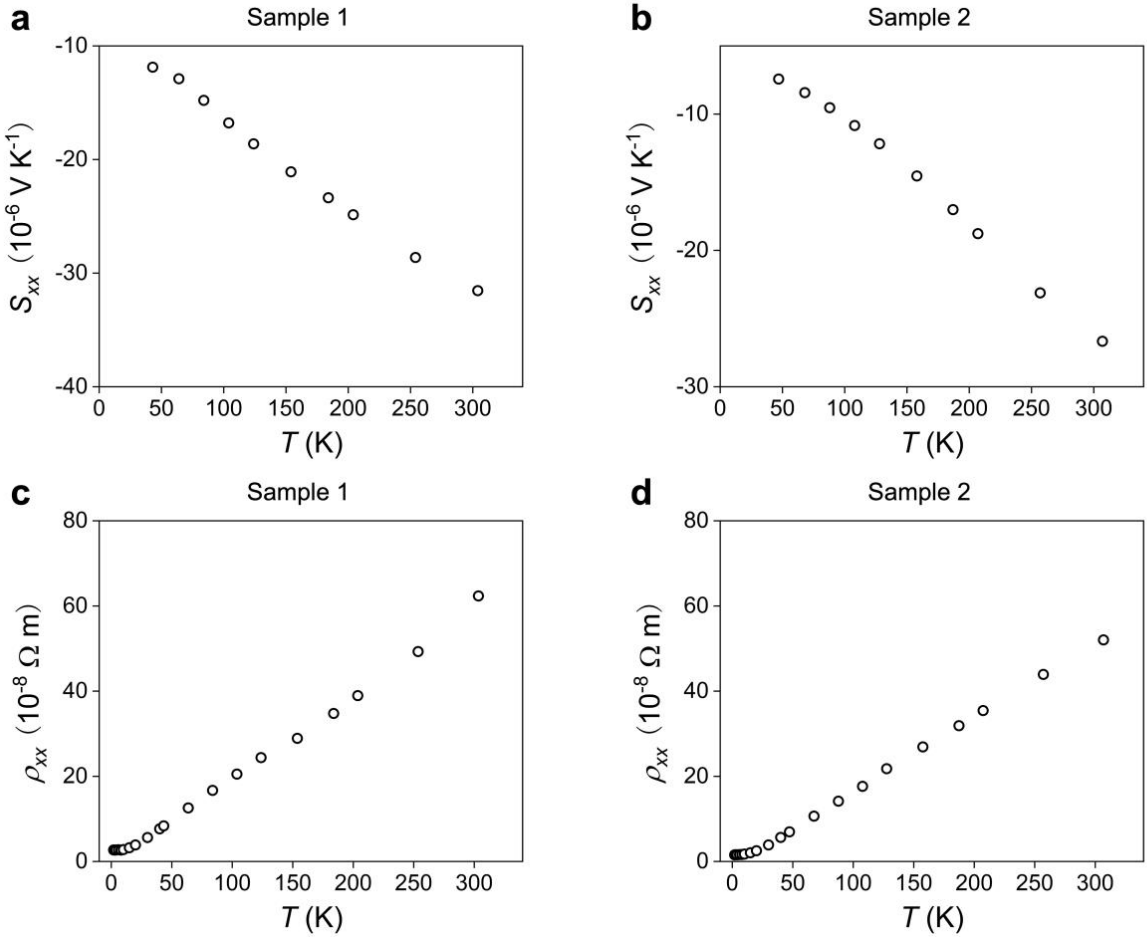

**Supplementary Fig. 11 | Temperature dependence of Seebeck coefficient $S_{xx}$ and electrical resistivity $\rho_{xx}$ of samples 1 and 2**. **a**–**b**, Temperature dependence of Seebeck coefficient $S_{xx}$ of sample 1(**a**) and sample 2 (**b**). **c**–**d**, Temperature dependence of electrical resistivity $\rho_{xx}$ of sample 1(**c**) and sample 2 (**d**).

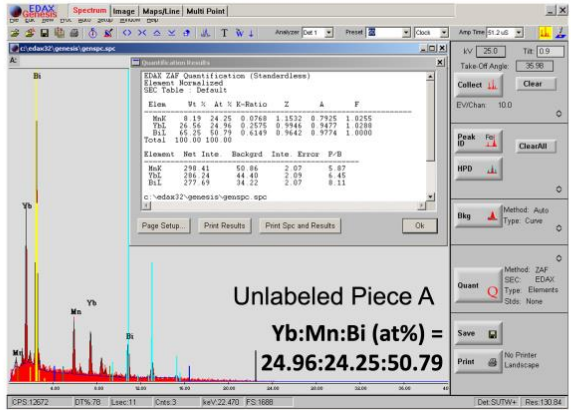 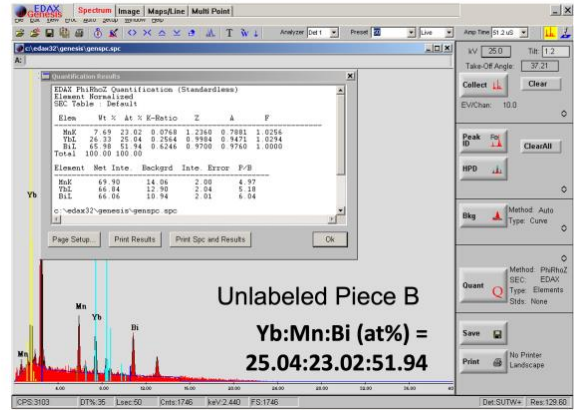

**Supplementary Fig. 12 | In-depth EDX analysis on other grown single crystals**. No significant compositional non-uniformities are observed. The maximum element variation (the column labelled K-ratio) observed is up to 3% within the limit of the experiment.

| Compound | Experimental Setup | Hall density (at 8 K) | Corresponding Hall mobility (at 8 K) | Maximum Total Nernst /ANE thermopower |
|---|---|---|---|---|
| YbMnBi$_2$ (sample 1 this work) | $H$ // [110] $j_Q$ // [1$\bar{1}$0] $V$ // [001] | $2.22 \times 10^{19}$ cm$^{-3}$ | $1.05 \times 10^5$ cm$^2$ V$^{-1}$ s$^{-1}$ | ~ 110 / > 70 µV K$^{-1}$ |
| YbMnBi$_2$ (sample 2 this work) | $H$ // [110] $j_Q$ // [1$\bar{1}$0] $V$ // [001] | $1.62 \times 10^{20}$ cm$^{-3}$ | $2.38 \times 10^4$ cm$^2$ V$^{-1}$ s$^{-1}$ | ~ 38 / ~ 30 µV K$^{-1}$ |
| YbMnBi$_2$ (ref. 17) | $H$ // [100] $j_Q$ // [010] $V$ // [001] | $1.95 \times 10^{20}$ cm$^{-3}$ | $6.49 \times 10^3$ cm$^2$ V$^{-1}$ s$^{-1}$ | ~ 10 / ~ 6 µV K$^{-1}$ |
| YbMnBi$_2$ (ref. 17) | $H$ // [100] $j_Q$ // [001] $V$ // [010] | $2.25 \times 10^{20}$ cm$^{-3}$ | $2.08 \times 10^2$ cm$^2$ V$^{-1}$ s$^{-1}$ | ~ 6 / ~ 3 µV K$^{-1}$ |

**Supplementary Table 1 | Comparison of the samples and maximum total or anomalous Nernst signals between this study and previous research.** Hall densities for samples in this work are determined from the slope of the Hall resistance in the field range -0.1 T < $\mu_0 H$ < 0.1 T. The Nernst thermopowers are given both for the total and for the anomalous contribution when possible to be extracted, which is limited to the range where the ONE is linear in field. This procedure does not reliably apply to sample 1 of the present work. An estimate of greater than 70 µV K$^{-1}$ is mentioned here based on the high-field saturation tendency values above 180 K, because in the high-field regime (the product of mobility times field being much larger than unity) the ordinary Nernst coefficient tends to zero.

| T (K) | Sample 1 | | Sample 2 | |
| --- | --- | --- | --- | --- |
| | $A_0$ (A T m$^{-1}$ K$^{-2}$) | $B_0$ (T) | $A_0$ (A T m$^{-1}$ K$^{-2}$) | $B_0$ (T) |
| 43 | 0.000242 | 0.4689 | 0.01560 | 5.5991 |
| 64 | 0.000313 | 0.4806 | 0.01070 | 5.2646 |
| 84 | 0.000444 | 0.5887 | 0.00750 | 4.8657 |
| 104 | 0.000673 | 0.8633 | 0.00522 | 4.3251 |
| 124 | 0.000827 | 1.0801 | 0.00373 | 3.8116 |
| 154 | 0.001068 | 1.5254 | 0.00251 | 3.3113 |
| 184 | 0.001300 | 2.1104 | 0.00196 | 3.2641 |
| 204 | 0.001412 | 2.5647 | 0.00173 | 3.3747 |
| 254 | 0.001528 | 3.9566 | 0.00144 | 4.2255 |
| 304 | 0.001420 | 5.6968 | 0.00126 | 5.7263 |

**Supplementary Table 2 | Fitting parameters for theoretical description of the Nernst conductivity.** The resulting fitting parameters $A_0$ and $B_0$ for describing the Nernst conductivity $\alpha_{yx}$ of both sample 1 and sample 2 using the theoretical equation (S22) described in the Supplementary Information.